
\input phyzzx
\hoffset=0.2truein
\hsize=6truein
\voffset=0.1truein
\def\TITLEPAGE{\frontpagetrue}
\def\IASSNS#1{\hbox to\hsize{\tenpoint \baselineskip=12pt
        \hfil\vtop{
        \hbox{\strut iassns-hep-94-#1}
}}}
\def\PUPT#1{\hbox to\hsize{\tenpoint \baselineskip=12pt
        \hfil\vtop{
        \hbox{\strut PUPT-94-#1}
}}}
\def\HEP{\hbox to\hsize{\tenpoint \baselineskip=12pt
        \hfil\vtop{
        \hbox{\strut hep-ph/9411206}
}}}
\def\SUNYSB{
\centerline{$^*$
 Institute for Theoretical Physics, State University of New York}
\centerline{Stony Brook, New York 11794-3840}}
\def\PRINCETON{
\centerline{$^\dagger$Physics Department, Princeton University}
\centerline{Princeton, New Jersey 08544}}
\def\IAS{
\centerline{$^+$School of Natural Science, Institute for Advanced Study}
\centerline{Olden Lane, Princeton, New Jersey 08540}}
\def\TITLE#1{\vskip .0in \centerline{\fourteenpoint #1}}
\def\AUTHOR#1{\vskip .1in \centerline{#1}}

\def\ABSTRACT#1{\vskip .1in \vfil \centerline{\twelvepoint
\bf Abstract}
   #1 \vfil}
\def\ENDTITLEPAGE{\vfil\eject\pageno=1}
\hfuzz=5pt
\tolerance=10000
\TITLEPAGE
\IASSNS{81}
\PUPT{1507}
\HEP
\TITLE{An Open Universe From Inflation}

\AUTHOR{Martin Bucher$^{\dagger,+}$}
\AUTHOR{Alfred S. Goldhaber$^*$}
\AUTHOR{Neil Turok$^\dagger$}
\IAS
\PRINCETON
\SUNYSB
\nobreak
\ABSTRACT{
We present a natural scenario
for obtaining an open universe ($\Omega _0<1$) through inflation. In this
scenario, there are two epochs of
inflationary expansion---an epoch of `old inflation,' during which the
inflaton field is stuck in a false vacuum, followed by an epoch of `new
inflation,' during which the inflaton field slowly rolls toward its true
minimum. During the first epoch,
inflation solves the smoothness and horizon
problems. Then an open universe (with negative spatial curvature) is created
by the nucleation of a single
bubble. In effect $\Omega$ is instantaneously `reset' to zero.
During the subsequent
`new' inflation $\Omega$ rises toward unity.
The value of $\Omega$ today is calculable in terms of the
parameters of the potential, and we show that obtaining values
significantly different from zero or unity (though within
the range $0<\Omega <1$) does not require significant fine tuning.
We compute the spectrum of density perturbations by evolving
the Bunch-Davies vacuum modes across the bubble wall into its interior.
}
\rightline{[October 1994---Revised: February 1995]}

\ENDTITLEPAGE

\eject

\REF\dolginov{ A.Z. Dolginov and I.N. Toptygin,
``Relativistic Spherical Functions,"
Zh. Eksperim. i. Teor. Fiz. {\bf 37,} 1441 (1959).
[Translation: Sov. Phys. JETP {\bf 10,}
1022 (1960).]}

\REF\lhm{
 A.S. Goncharov and A.D. Linde, ``
Tunneling in Expanding Universe: Euclidean and Hamiltonian Approaches'',
Sov. J. Part. Nucl. 17 ( 1986) 369;
A. Linde, Stochastic Approach to Tunneling an Baby Universe Formation''
 Nucl. Phys. B372 (1992) 421.}

\REF\abbott{L. Abbott and R. Schaefer, ``A General,
Gauge-Invariant Analysis of the
Cosmic Microwave Anisotropy," Ap. J. {\bf 308,} 462 (1986).}

\REF\mirror{R. Carlitz and R. Wiley, ``Reflections on Moving
Mirrors," Phys. Rev. {\bf D36,} 2327 (1987).}

\REF\birrell{ N. Birrell and P. Davies, {\it Quantum Fields in
Curved Space,} (Cambridge, Cambridge U. Press, 1982) and
references therein.}

\REF\gott{J.R. Gott, III, ``Creation of Open Universes from de Sitter Space,"
Nature {\bf 295,} 304 (1982); J.R. Gott and T. Statler, ``Constraints on the
Formation of Bubble Universes," Phys. Lett. {\bf 136B,} 157 (1984);
J.R. Gott, ``Conditions for the Formation of Bubble Universes," in
E.W. Kolb et al., Eds., {\it Inner Space/Outer Space,} (Chicago:
U. of Chicago Press, 1986).}

\REF\lyth{D. Lyth and E. Stewart, ``Inflationary Density Perturbations
with $\Omega <1,$'' Phys. Lett. {\bf B252,} 336 (1990).}

\REF\rp{B. Ratra and P.J.E. Peebles, ``Inflation in an Open Universe,"
PUPT-1444 (Feb. 1994).}

\REF\rptwo{B. Ratra and P.J.E. Peebles,
``CDM Cosmogony in an Open Universe," Ap. J. Lett. {\bf 432}, L5 (1994).}

\REF\rub{V. Rubakov, ``Particle Creation During Vacuum Decay,"
Nucl. Phys. {\bf B245,} 481 (1984).}

\REF\vv{T. Vachaspati and A. Vilenkin,
``Quantum State of a Nucleating Bubble,"
Phys. Rev. {\bf D43,} 3846 (1991);
J. Garriga and A. Vilenkin, `Quantum Fluctuations on Domain Walls,
Strings, and Vacuum Bubbles," Phys. Rev. {\bf D45,} 3469 (1992).}

\REF\sasaki{ M. Sasaki, T. Tanaka, K. Yamamoto, and J. Yokoyama,
``Quantum State During and After Nucleation of an
$O(4)$ Symmetric Bubble,";  Prog. Theor. Phys. {\bf 90,}
1019 (1993); M. Sasaki, T. Tanaka, K. Yamamoto, and J. Yokoyama,
``Quantum State Inside a Vacuum Bubble and Creation of
an Open Universe," Phys. Lett. {\bf B317,} 510 (1993.}

\REF\sasakinew{  T. Tanaka and M. Sasaki,
``Quantum State During and After O(4) Symmetric Bubble
Nucleation with Gravitational Effects", Phys. Rev. {\bf D50,}
6444 (1994).}

\REF\gh{G. Gibbons and S. Hawking,
``Cosmological Event Horizons, Thermodynamics,
and Particle Creation," Phys. Rev. {\bf D15,} 2738 (1976).}

\REF\ba{B. Allen, ``Vacuum States in de Sitter Space," Phys.
Rev. {\bf D32,} 3136 (1985).}

\REF\bv{A. Vilenkin, ``Did the Universe Have a Beginning?",
Phys. Rev. {\bf D46,} 3255 (1992);
A. Borde and A. Vilenkin, ``Eternal Inflation and the Initial
Singularity," Phys. Rev. Lett. {\bf 72,} 3305 (1994).}

\REF\bander{M. Bander and C. Itzykson, ``Group Theory and
the Hydrogen Atom (II)," Rev. Mod. Phys. {\bf 38,} 346 (1966).}

\REF\cc{S. Coleman, Phys. Rev. {\bf D15,} 2929 (1977), {\bf D16,}
1248(E) (1977);
C. Callan and S. Coleman, Phys. Rev. {\bf D16,} 1763 (1977); see also,
``The Uses on Instantons," in S. Coleman, {\it Aspects of Symmetry,}
(Cambridge, Cambridge University Press, 1985).}

\REF\cd{S. Coleman and F. De Luccia, ``Gravitational Effects on and of
Vacuum Decay," Phys. Rev. {\bf D21,} 3305 (1980).}

\REF\guth{A. Guth, ``Inflationary Universe: A Possible Solution to the Horizon
and Flatness Problems," Phys. Rev {\bf D23,} 347 (1981).}

\REF\gw{A. Guth and E. Weinberg, ``Could the Universe Have Recovered from a
First-Order Phase Transition?," Nucl. Phys. {\bf B212,} 321 (1983).}

\REF\lninf{
A. Linde, ``A New Inflationary Universe Scenario: A Possible Solution of the
Horizon, Flatness, Homogeneity, Isotropy, and Primordial Monopole Problems,"
Phys. Lett. {\bf 108B,} 389 (1982).}

\REF\aninf{A. Albrecht and P. Steinhardt, ``Cosmology for Grand Unified
Theories
with Radiatively Induced Symmetry Breaking," Phys. Rev. Lett. {\bf 48,} 1220
(1982).}

\REF\linch{
A. Linde, ``Chaotic Inflation'',
Phys. Lett. {\bf 129B,} 177 (1983).}

\REF\jensen{L. Jensen and P. Steinhardt, ``Bubble Nucleation and the
Coleman-Weinberg Model," Nucl. Phys. {\bf B237,} 176 (1984).}

\REF\iperth{S. Hawking, ``The Development of Irregularities in a Single
 Bubble Inflationary Universe'', Phys. Lett. {\bf 115B}, 295 (1982).}

\REF\iperts{A.A. Starobinsky, ``Dynamics of Phase Transition in the
New Inflationary Scenario and Generation of Perturbations'',
  Phys. Lett. {\bf 117B}, 175 (1982).}

\REF\ipertgp{A.H. Guth and S.-Y. Pi, ``Fluctuations in the New Inflationary
Universe'',
 Phys. Rev. Lett. {\bf 49}, 1110 (1982).}

\REF\ipert{J. Bardeen, P. Steinhardt, and M. Turner, ``Spontaneous Creation of
Almost Scale-Free Density Perturbations in an Inflationary Universe,"
Phys. Rev. {\bf D28,} 679 (1983).}

\REF\kodama{H. Kodama and M. Sasaki, ``Cosmological Perturbation
Theory," Prog. of Theor. Phys. Suppl. {\bf 78,} 1 (1984).}

\REF\mfb{V. Mukhanov, H. Feldman and R. Brandenberger, ``Theory of Cosmological
Perturbations," Phys. Rep. {\bf 215,} 203 (1992).}

\REF\bardeen{J. Bardeen, ``Gauge Invariant Cosmological
Perturbations," Phys. Rev. {\bf D22,} 1882 (1980).}

\REF\nextpaper{M. Bucher and N. Turok, ``Large Angle CMB Anisotropy
in an Inflationary $\Omega <1$ Universe," in preparation.}

\REF\allencaldwell{B. Allen and R. Caldwell, ``Cosmic Background
Radiation Temperature Fluctuations in a Spatially Open
Inflationary Universe'', unpublished manuscript
(1994).}

\REF\sasakitanaka{ K. Yamamoto, T. Tanaka and M. Sasaki,
``Particle Creation through Bubble Nucleation and Quantum Field Theory
in the Milne Universe'', preprint in preparation (1994).}

\REF\lythtwo{D. Lyth, ``Large Scale Energy Density Perturbations
and Inflation," Phys. Rev. {\bf D31,} 1792 (1985).}

\REF\btnew{M. Bucher and N.Turok, ``Open Inflation with
Arbitrary False Vacuum Mass," Princeton Preprint PUPT 95-1518
(2-95).}

\REF\yamnew{K. Yamamoto, M. Sasaki and T. Tanaka,
``Large Scale CMB Anisotropy in an Open Universe in the
One-Bubble Inflationary Scenario'', preprint KUNS 1309,
astro-ph/9501109 (1995).}

\def\Box{\mathord{\dalemb{5.9}{6}\hbox{\hskip1pt}}}
\def\dalemb#1#2{{\vbox{\hrule height.#2pt
	\hbox{\vrule width.#2pt height#1pt \kern#1pt \vrule width.#2pt}
	\hrule height.#2pt}}}
\def\kh{k_{h}}
\def\zh{\zeta _{h}}
\def\zd{\zeta _{dS}}
\def\zd{{\zeta _{dS}}}
\def\Box{\mathord{\dalemb{5.9}{6}\hbox{\hskip1pt}}}
\def\dalemb#1#2{{\vbox{\hrule height.#2pt
	\hbox{\vrule width.#2pt height#1pt \kern#1pt \vrule width.#2pt}
	\hrule height.#2pt}}}
\def\sech{ {~\rm sech ~ }}

\def\gtorder{\mathrel{\raise.3ex\hbox{$>$}\mkern-14mu
             \lower0.6ex\hbox{$\sim$}}}
\def\ltorder{\mathrel{\raise.3ex\hbox{$<$}\mkern-14mu
             \lower0.6ex\hbox{$\sim$}}}

\def\overleftrightarrow#1{\vbox{\ialign{##\crcr
$\leftrightarrow$\crcr\noalign{\kern-1pt\nointerlineskip}
$\hfil\displaystyle{#1}\hfil$\crcr}}}
\def\dpartial{{\overleftrightarrow{\partial}}}

\chapter{Introduction}

According to conventional lore, inflation\refmark{\guth} predicts a
spatially flat universe with the
cosmological density parameter $\Omega _0$ being very close
to unity. It is
therefore interesting to inquire whether inflation would
be ruled out as a viable cosmological theory,
or at least
rendered sufficiently contrived to lose much of its underlying
beauty and attractiveness, if it were to become observationally
established
that we live in an open universe with $\Omega _0<1$ today.
In this paper we present a class of inflationary models
that generate such an open universe in a natural way.
This means that observations indicating $\Omega_0$ is
less than unity can only limit the class of
possible inflationary models and {\it cannot} rule out the whole idea.
Our calculations shall show that in a sense the
inflationary scenario is compatible with any value of
$\Omega_0$ between zero and unity.

Before presenting our scenario we first
review the standard non-inflationary
 argument against $\Omega _0\ne 1.$
For an equation of state of the form $\rho \sim a^{-\gamma },$
$(\Omega^{-1} -1)$ scales as $a^{\gamma -2}$,
where $a$ is the scale factor of the universe.
This means that during a radiation dominated or matter dominated
epoch, $\Omega $ flows away from one. However,
during an inflationary epoch (defined here by
the condition $\gamma <2$) $\Omega $ flows toward one.
In the usual computation
of the extreme unnaturalness of $\Omega_0$ being
very different from unity, one
evolves $\Omega$ back
to some much earlier time assuming that $\gamma \approx 3$--$4$ and notes
just how precisely $\Omega $ must coincide with unity at that early
epoch in order
to evolve into a value of the order of unity today.
If $\Omega_0$ is not exactly one,
the deviation of this earlier value of $\Omega $ from one gives a
very small number, which is taken to reflect the degree of fine
tuning required to obtain a nonflat universe.

Such a calculation, however,
is inappropriate if there is a preceding inflationary epoch, during which
co-moving scales are  pushed out of the Hubble radius, because in such a
scenario the co-moving scale $x_{curv}$ characterizing the curvature of an
open (or closed) universe is within the Hubble radius sometime during
inflation,
or perhaps at the beginning of inflation. If curvature is
introduced as an initial condition prior to inflation, inflation acts
to mitigate some of the unnaturalness of $\Omega \ne 1,$ but at a high
price. By tuning the length of inflation to be short, one can  by
imposing curved
initial conditions arrange to have significant curvature
today, but in this case one would also have to put in
by hand smoothness on scales of order
$x_{curv}$, and thus much of
the initial motivation for proposing inflation would be lost.

The natural way to obtain an open universe from inflation is to introduce
spatial curvature {\it during} inflation, so that whatever inhomogeneities
may exist prior to
inflation are still erased, but spatial curvature is regenerated.
In this way, inflation still provides a solution to the
smoothness and horizon problems, while producing a nonflat universe. In such a
scenario little
 fine tuning is required, because it is the length of inflation {\it
after}
the event introducing the curvature that must be adjusted to obtain the proper
$\Omega _0.$ Said another way, one does not tune $\Omega _0$ but rather the
logarithm of its difference from one, which requires little fine
tuning.

The remarkable fact that one can introduce spatial curvature
within an inflating universe in a completely causal way, and
without spoiling its homogeneity, was first revealed in the
calculations of Coleman and De Luccia \refmark{\cd}, prior to the
invention of inflation. Their calculations showed that
the spacetime interior of a (finite) nucleating bubble actually
encompasses an (infinite) open FRW universe.
The idea that a realistic open universe could emerge from de Sitter
space was first proposed by Gott in 1982,\refmark{\gott }
prior to the
introduction of the new inflationary scenario, and
independently of Coleman and De Luccia's work.

A realistic open universe scenario may be obtained via this mechanism,
by combining {\it old inflation}
with {\it new inflation.} \footnote{\dagger}{
We emphasize here that what we mean
by new inflation is simply inflation that  proceeds whilst
a scalar field is rolling slowly down a potential (i.e.
`slow-roll' inflation). In its original incarnation, the
new inflationary scenario \refmark{\lninf, \aninf}
assumed the universe began in
a state close to thermal equilibrium before inflation began---%
this was presumed to explain how the inflaton field became
localized around a potential maximum. We shall make the much milder
assumption that {\it somehow} a scalar field became trapped
in a false vacuum over a sufficiently large region for
inflation to begin. This could have happened as a result of
`random' initial conditions, in which case our scenario
would be more akin to
chaotic inflation, \refmark{\linch} `special' initial conditions, such as
the Hartle-Hawking-Vilenkin proposals, or even through
a state of thermal equilibrium. Once the scalar field
becomes trapped, and the epoch of
`old' inflation begins, the {\it details} of how it became trapped
are of course very quickly erased. }
 The universe begins in a false
vacuum in what might be described as the `old inflationary epoch,'
during which any pre-existing inhomogeneities
are redshifted away.
The smoothness and horizon
problems are solved during this first epoch of old inflation.
Then a single bubble nucleates, inside of which the entire presently
observable universe is formed.  After
bubble materialization, the surfaces on which the inflaton field $\phi ,$
which we shall for the most part
assume to be a single real scalar field, is constant are
surfaces of constant negative spatial curvature.
[If one ignores gravitational effects
and thus considers bubble nucleation in Minkowski space,
$\phi $ is constant on surfaces of constant $s^2=(t^2-x^2-y^2-z^2),$
which for $s^2>0$ are surfaces of constant negative spatial
curvature.] It is these surfaces that will become surfaces
of constant cosmic density, or constant cosmic `time' in
the usual sense.
However, if one would proceed
directly to an FRW
post-inflationary epoch by reheating near the bubble wall one would
obtain
$\Omega $ somewhat less than one at re-heating, and the present value
$\Omega _0$ would be
extremely close to zero. Such a scenario is clearly unacceptable. Instead we
propose passing to an
epoch of new inflation after bubble nucleation, tunneling from the false vacuum
onto a slow-rolling potential. To obtain an open universe one wants the
Hubble radius shortly after bubble nucleation to be of
approximately the same co-moving size as
the present Hubble radius, so that $\Omega _0$ is near one.
This involves adjusting the length of the epoch of new inflation. In models
of new inflation designed to explain $\Omega _0=1$
the length of new inflation is arbitrary, provided that
it is long enough to erase the initial inhomogeneities. What happens at the
end of inflation is observationally accessible, but the earlier part
of inflation is not,
because what happened within the Hubble radius at earlier times now lies hidden
on co-moving scales much larger than the Hubble radius of our
presently observable universe.
In the inflationary scenario that we consider, the entire new inflationary
epoch is observationally accessible, and part of the old inflationary
epoch as well.

We emphasize that none of the ideas we have discussed so far are
in any way new. Indeed they follow precisely the historical
lines of development of `old' and `new' inflation.
However we believe that the predictive nature
of the `single-bubble' scenario
in a situation where the spatial curvature inside the bubble
is {\it not} completely wiped out by `new' inflation
inside the bubble has been overlooked.
It is this significant
lacuna in the literature which this paper seeks to fill.

The bulk of this paper deals with calculating the spectrum
of density perturbations in the open inflationary model
just described. On `small' co-moving scales---where small
is defined relative to the scale of spatial curvature---the
spectrum of density perturbations is approximately scale
invariant, just as in conventional `new' inflation. This is
because small scales leave the Hubble radius  at late times during
the epoch of new inflation, when the curvature scale
is much larger than the Hubble radius. Hence curvature is
of little relevance in determining the small-scale
perturbations.

By contrast, on larger co-moving scales scales negative
spatial curvature becomes increasingly relevant,
and we expect deviations from scale invariance to occur.
For large scales quantum fluctuations
that arise during the epoch of old inflation are also
relevant. There are also other effects from the change
in $\partial ^2V/\partial \phi ^2$ on passing through the
bubble wall that lead to parametric excitations. These may be called
`moving mirror' effects \refmark{\birrell , \mirror }.

The zeroth order classical expanding bubble
solution is $SO(3,1)$ invariant and described by
a coordinate system that divides spacetime into three
regions, as indicated in Figure 1. In region I the metric
has the form
$$ds^2=-dt^2+a^2(t)\cdot \bigl[ d\xi ^2+\sinh ^2[\xi ]~
d\Omega _{(2)}^2\bigr]
\eqn\metricaa$$
where $ d\Omega _{(2)}^2=d\theta ^2+\sin ^2\theta d\phi ^2.$
The metric \metricaa ~ represents an open expanding FRW
universe. An important feature of this metric is that
it is nonsingular - the usual `big-bang' singularity
at $a(t) =0$ is here just a coordinate singularity - at small
times $a(t) \sim t$ and the metric [\metricaa ] describes the
Milne universe (see e.g. ref. \birrell), just a coordinate rewriting of
Minkowski space.
In region II the metric has the form
$$ds^2=d\sigma ^2+b^2(\sigma )\cdot
\bigl[ -d\tau ^2+\cosh ^2[\tau ]~d\Omega _{(2)}^2\bigr]
\eqn\metricab$$
and $b(\sigma )/\sigma \to 1$ as $\sigma \to 0.$
In region III, which essentially is an open FRW universe
expanding backward in time, the metric the same form as in eqn.
\metricaa . In the perfect expanding bubble solution
the scalar field is constant on surfaces of constant $t$
and constant $\sigma $ as well as on the light cone of
the origin, which separates the three regions.

Only the upper half of the classical solution,
sketched in in Figure 1, is physically realized. The lower half, which
represents a bubble expanding backward in time, should be
replaced with an instanton, known as the Coleman--de Luccia
instanton, representing the classically forbidden process
of bubble nucleation, as sketched in Figure 2. The instanton
and the classical expanding bubble solutions are trivially related
to each other by analytic continuation.

At first it may seem paradoxical that an infinite open universe
can fit inside an expanding bubble, which at any
moment would seem to have a increasing finite (but never
infinite) size, increasing at approximately the speed of
light. Indeed if one were to use a time foliation in figure 2
in which horizontal lines would correspond to constant time
hypersurfaces, at any finite time the bubble would have a finite
size, and an observer using such a coordinate system would be
quite justified in saying that the size of the bubble is
always finite. But such a foliation would not correspond to what
we today would consider a natural definition of constant time.
For us constant cosmic time corresponds to a constant CMB temperature.
To us a foliation that makes the spatial homogeneity and isotropy
of the universe manifest seems preferred, and such a time
slicing corresponds to the hyperboloids skteched in region I
of figure 2, which are infinite. This is how an infinite
open universe can fit inside a bubble. Despite their
contradictory appearance, the two points of view are
quite compatible with each other.

Another possible source of density perturbations, which we do not
consider here, arises from imperfections in the initial bubble.
In the semiclassical limit $\hbar \to 0$ the scalar field
configuration that results from quantum tunneling is fixed,
with no random variations from the perfect expanding bubble
solution. But as one considers corrections of
higher order in $\hbar$, small random variations occur, because
the field does not tunnel exactly along the configuration of
least Euclidean action, but rather along a nearby path. We
believe that such perturbations in the tunneling
process itself are insignificant, for two reasons.
First, if the bubble nucleation is
a strongly suppressed process, the fluctuations
around it are likely to be even more strongly suppressed.
And second, because in the
subsequent classical expansion of the bubble
the details of how the bubble was initially formed
are likely to be quickly erased. The dominant sources of perturbations
rather are the quantum fluctuations of the fields in
de Sitter space prior to
bubble nucleation, which pass across the bubble wall
into its interior, and those that evolve later inside the bubble.

Open inflation
has been considered more recently
in refs. \lyth -\rptwo . Lyth and Stewart,\refmark{\lyth} and
Ratra and Peebles\refmark{\rp, \rptwo }
have considered density perturbations
in an open inflationary scenario, a scenario of new inflation
with negative spatial curvature put in by hand. They do
not base their calculation of the spectrum of density
perturbations on a specific model for the creation of negative
curvature at the beginning of their inflationary scenario. Rather
the discussion is confined to what we call region I, and as
initial conditions for the fluctuations of the inflaton field
at $t=0,$
they impose the requirement that operators corresponding
to the modes with asymptotic behavior $t^{-i\zeta -1}$ (as opposed to
$t^{+i\zeta -1}$) as $t\to 0+$ annihilate the vacuum.
This state is sometimes referred to as the `conformal vacuum',
and it has the following unappealing property.
The energy density in scalar field
fluctuations in their assumed quantum
state actually diverges as one approaches $t \to 0+$, which means
that the coordinate singularity in the background solution at $t=0$
becomes a {\it real} singularity in this state.
Thus one loses one of the principal attractions of open inflation,
that the standard big bang singularity is removed.
In contrast,
with our choice of
initial state, the energy density is
nonsingular as $t \to 0+$,
and continues smoothly through the coordinate singularity at $t=0$,
into  de Sitter space.
So whereas Lyth and Stewart, and
Ratra and Peebles,  have, for consistency, to simply assume
their universe begins on a spatially homogeneous and isotropic
slice some short time after $t=0$, and thus give up on the inflationary
solution of the horizon/homogeneity puzzles, we have as our starting
point a long era of `old' inflation outside the single-bubble
universe, which in which those problems are solved.

The quantum state of a scalar field of constant mass
outside an expanding bubble in Minkowski space
was considered by Rubakov,\refmark{\rub }, by
Vachaspati and
Vilenkin, and Garriga and Vilenkin\refmark{\vv},
using the thin-wall approximation.
Sasaki et al. \refmark{\sasaki } considered the quantum
state inside such a bubble in the usual thin-wall approximation.
Recently they have extended this work to the case
of de Sitter space, but where the interior of the bubble is
empty, so no `new' inflation and reheating occurs.\refmark{\sasakinew }

The organization of the paper is the following.
In section II we give a simple discussion
of how the value of $\Omega_0$ depends
on the parameters of the inflaton potential $V(\phi)$
and the reheating temperature.
In section III
we give a  brief review of bubble nucleation in de Sitter
space in the semiclassical limit, in which there are no
perturbations. The Euclidean bounce is the zeroth order
solution, about which we consider small perturbations
treated as free quantum fields.
 Section IV contains
a new `thin wall' approximate bounce solution which
we use to show that there is a well defined
region of parameter space in which significant simplifications
of the nucleating and expanding bubble solutions occur.
 The analytic continuation
of the Euclidean bounce into a spacetime with a Lorentzian
spacetime signature possesses an $SO(3,1)$ symmetry
which we exploit
to simply our calculations.
Section V
develops the formalism necessary to do  this.
In section VI we consider initial conditions
prior to bubble nucleation. This involves rewriting the
two-point function for the Bunch-Davies vacuum in terms of
the region II mode functions. In section VII we deal with
the evolution of cosmological perturbations in region I.
In section VIII we deal with re-heating and observational
consequences, and finally in section IX we present some
concluding remarks.

\chapter{Calculating $\Omega$ from Inflation}

The scenario we shall explore in this paper is based
on a scalar potential $V(\phi)$ of the general
form illustrated Figure 3.
We assume that the scalar field $\phi$ became stuck
in a `false vacuum' F in a region of the universe large
enough for an extended period of inflation to occur.
Once this has happened, it is very plausible that the universe
is to a very good approximation
described as de Sitter spacetime. This state of affairs
is interrupted by the nucleation of a bubble.
In a localized region of space
the inflaton field tunnels quantum mechanically through the
potential barrier from its initial value $\phi_i$ to
a value $\phi_n$, the value of the
field at the center of the nucleated bubble
at the instant of nucleation. We describe the nucleation process
semiclassically, using a Euclidean `bounce' solution.
Inside the bubble, the field $\phi$ rolls down the potential
toward $\phi_r$, where oscillations of the scalar field
reheat the universe and lead to a conventional radiation
dominated universe \footnote\dagger{The detailed structure of $V(\phi)$
around $\phi_r$ shall play little role in our discussion.}
A spacetime picture of the bubble
nucleation and subsequent reheating is shown in Figure 2.
$\phi = \phi_n$ on the future light cone of
the point at the center of the nucleated bubble.
Surfaces of constant $\phi$ within this light cone are
spaces of constant negative curvature (open universes).
They are also surfaces of constant density in
the subsequent evolution of the universe, which we would call
surfaces of constant cosmic time today. But these surfaces are
of course infinite. So inside
a finite bubble one has produced an infinite open universe.
This remarkable picture was first revealed in the calculations
of Coleman and de Luccia. \refmark{\cd}

We shall begin with a simple point.
In  this scenario
the initial conditions prior to bubble nucleation
and `new' inflation are very  definite---the universe
has become very homogeneous and the value of $\phi$ has been
fixed. From this follows a great deal of predictive power. In particular
the value of $\Omega$ today is not a free parameter; it is
fully determined by the form of the potential, and by the microphysical
details of the reheating process. So one can sensibly ask the question of how
much `fine tuning' of the parameters in the
potential is needed in order to
produce a given value of $\Omega $ today.
This is easily calculable with
some reasonable simplifying approximations.
We shall
assume that the potential well around
$\phi_i$ is narrow, so that $\phi_n \approx \phi_i$.
We further  assume that for most of the region
between $\phi_n$ and $\phi_r$,  the linear approximation
$$
V(\phi) \approx - \mu^3 \phi
\eqn\naa
$$
is valid.
(This is easily generalized to an arbitrary power
law form for $V(\phi)$, with very minor
changes in the conclusions).
For simplicity we have
defined the value of $\phi$ at reheating,
$\phi_r$ in Figure 3, to be  zero. As in all inflationary scenarios,
we have to adjust the potential so that $V=0$ (no cosmological constant)
in the universe after reheating.
The motion of $\phi$ is described by the equation
$$
\ddot{\phi} + 3 H\dot{\phi}= -{\partial V\over \partial \phi}
\eqn\na
$$
with $H$ as usual the Hubble constant, and we
take $\phi$ to be homogeneous on surfaces of constant
cosmic time $t$. The Einstein equation in an open universe
is just
$$
H^2 =\left( {\dot{a} \over a}\right) ^2
 = {8 \pi G \over 3} ({1\over 2} \dot{\phi}^2 + V(\phi))  + {1\over a^2}
\eqn\nb
$$
where $a(t)$ is the scale factor. These equations describe
the classical field solution for the interior of
a nucleating bubble
(see Figure 2).

In the `slow-roll' approximation, which becomes very good
after a short initial transient, the $\ddot{\phi}$ term in
(\na ) and the $\dot{\phi}^2$ term in (\nb ) may be dropped.
The scale factor $a(t)$ is then well described by an
adiabatic  approximate solution to \nb:
$$
a(t) \approx {1\over B} ~
{\rm sinh} \left[ \int_0^t dt~B \right] ,\qquad \qquad
B^2 ={8 \pi G V(\phi) \over 3}.
\eqn\nc
$$
At early times, the spatial curvature dominates and
$a(t)\propto t$, but
as inflation sets in  $a(t)$ starts growing exponentially,
with $a(t)\propto e^{\int H dt}$.
Once this happens,
the curvature term in eqn. \nb ~ becomes negligible  and
$H\approx B$. The number of e-foldings
during inflation is approximately
$$\eqalign{
\int dt~H &= \int H{d \phi \over \dot \phi} \approx  -  3
\int
{d \phi ~H^2\over V_{,\phi}(\phi)}\cr
& \approx -8\pi G \int
{d\phi ~V(\phi) \over V_{,\phi}(\phi)} = 4 \pi G ~
\Delta \phi^2 .\cr
}\eqn\nd$$

Thus the number e-foldings  depends only
on the total change in the scalar field  during
inflation, and not on the parameter determining the
slope of the potential, which we call $\mu$. (The magnitude of
the density perturbations produced during inflation
{\it does } depend on $\mu$, and requiring them to be small
requires that $\mu \ll m_{pl}$).

In this scenario
the value of $\Omega$ begins at zero on the null surface defined by
$t=0$
(or $\sigma =0$)
but rapidly approaches unity during the inflation
inside the bubble.
{}From eqn. \nb\ one has
$$
\Omega^{-1} \approx  {3 H^2 \over 8 \pi G V } = 1 + {1\over a^2 B^2}
\approx 1+ 4 e^{- {8 \pi G \Delta (\phi^2)} }
\eqn\ne
$$
where the last expression holds at reheating, and follows from eqn. \nd .
After reheating,
we have $\Omega^{-1} -1 \propto (\rho a^2)^{-1}$,
growing as $a^2$ in the radiation and as $a$ in the matter eras,
respectively. The value of $\Omega$ today is then given by
$$\Omega_0
\approx {1\over 1+ 4 e^{- {8 \pi G \Delta \phi^2}} {\cal A} };
\qquad \qquad {\cal A} \approx
\left( {T_{rh}\over
T_{eq}}\right) ^2 {T_{eq} \over T_0}
\eqn\nnf
$$
where $T_0$ is the $CMB$ temperature today, $T_{rh}$ that
after reheating and $T_{eq}$ that at equal density of matter and
radiation.\footnote\dagger{
We have here assumed instantaneous reheating. If reheating is slower,
there can be an intermediate, matter dominated
stage during which $\Omega^{-1}-1$ scales
inversely with temperature rather than temperature squared.
This is a minor numerical detail.}
If $T_r$ is of order the electroweak scale, $\approx 100$ GeV, then
${\cal A} \approx 10^{25}$, but if $T_{rh}$ is of order
the GUT scale, then ${\cal A} \approx 10^{50}$.
The qualitative behavior of eqn. \nnf ~ is easily understood.
If $\Delta \phi$ is small, then little inflation
occurs and $\Omega_0$ is very close to zero.
Conversely, if $\Delta \phi$ is large, a lot of inflation
occurs, and $\Omega_0 $ is very close to unity.
Generalizing the calculation to arbitrary power law form
for the potential, $V(\phi) \propto \phi^n$, one finds
that the factor $8 \pi$ is replaced by $8 \pi /n$.

Defining the value of $\phi$ at reheating to be zero,
the value of $\phi_i$ (or equivalently of
$\phi_n,$ because we have assumed they are nearly equal)
required to obtain $\Omega_0$ today
is given by
$$\phi_i^2 =
 {m_{pl}^2 n \over 8 \pi } ~{\rm ln}\left[
 {{\cal A} \over \Omega_0^{-1} -1}\right] .
\eqn\ne$$
As a concrete numerical example, let us assume reheating up to
the electroweak scale, so that
${\cal A} = 10^{25},$ and a linear potential, so that $n=1$.
A value of  $\Omega_0$ between .1 and .9 today then  requires
$1.48\ltorder (\phi_i/m_{pl})\ltorder 1.54 .$

Following this exercise, it seems to us that it is hard to argue
that inflation is incompatible with an
open universe.
In any fundamental theory, one expects
that the value of $\phi_i$ would be  fixed, presumably
in Planck mass units. It would not seem very implausible
that $\phi_i$ should lie within the few percent
of parameter space required to produce an interesting value of
$\Omega_0$ significantly different from unity today.

\chapter{Background Solution}

The details of bubble nucleation in a curved
space of nonvanishing constant curvature (i.e. de Sitter
space) were first worked out by Coleman and de Luccia,
\refmark{\cd }
who generalized earlier work on bubble nucleation in
flat Minkowski space,\refmark{\cc } where gravitational
back reaction from the nucleating bubble is not taken into
account.

Before outlining the modifications necessary to
include the effects of gravity,
we  give a lightning review of bubble nucleation
in Minkowski space through quantum tunneling at
zero temperature. The bubble nucleation rate is
$$\Gamma =A~\exp \left[ -{1\over \hbar }
S_E\left[ \phi _{b}(x)
\right] \right] \eqn\qqa$$
where $S_E$ is the Euclidean action and
$\phi _{b}(x)$ is the $SO(4)$ symmetric
Euclidean {\it bounce} solution, satisfying the
equation
$$
{\partial ^2\phi _{b}\over \partial s^2}
+{3\over s}~{\partial \phi _{b}\over \partial s}
-{\partial V\over \partial \phi }=0,
\eqn\qqb$$
where $s=(x^2+y^2+z^2+t^2)^{1/2}.$ The constant
$A$ has dimensions of $(mass)^4$, and is harder to calculate,
but may be estimated crudely
as $m^4$ where $m$ is the mass
of the scalar field in the false vacuum.

The Euclidean bounce solution should be interpreted
in the following manner. To materialize a bubble, the
scalar field must pass through a classically forbidden
region, under a classical potential barrier. For the Euclidean
bounce given above (with its materialization center
located at ${\bf x}=t=0$), the classically forbidden
evolution occurs during $t<0,$ and is described using
the Euclidean path integral formalism. The classical
bounce represents the path of least Euclidean action
to the classical turning point. For $t>0,$ the evolution
is classically allowed, and thus treated classically.
Formally one changes the signature of the
metric from Euclidean for $t<0$ to Lorentzian for $t>0.$
In the Minkowski space case, one can regard this
as distorting the path integral over $\phi$ into
the complex $\phi$ plane, to run over a (complex) saddle point.

We now review the generalization with gravity taken
into account. When $V[\phi _{fv}]>0,$ the underlying
spacetime is de Sitter space, whose
Euclideanized version is $S^4$
because under analytic continuation the metric
$ds^2=-dt^2+\cosh ^2[t]~d\Omega _{(3)}^2$
becomes
$ds^2=+dt^2+\cos ^2[t]~d\Omega _{(3)}^2,$ which
describes a four-dimensional sphere of radius
$R=H^{-1}.$

The equations of motion for the Coleman--de Luccia
bounce are derived by considering the Euclidean action
for the scalar field coupled to gravity, where
$SO(4)$ symmetry is imposed to restrict the form of the
metric to
$$ ds^2=d\sigma ^2+b^2(\sigma )~d\Omega _{(3)}^2 .\eqn\bnca$$
They are
$$\eqalign{
&\phi ^{\prime \prime }(\sigma )+{3b^\prime (\sigma )
\over b(\sigma )}~\phi ^\prime (\sigma )=
{\partial V\over \partial \phi },\cr
&\left[ {b'(\sigma )\over b(\sigma )}\right] ^2=
{1\over b^2(\sigma ) } + {8\pi G\over 3} \left\{
{1\over 2} \phi ^{\prime ~2}-V[\phi (\sigma )]
\right\} }\eqn\bncb$$
subject to the boundary conditions
$$\eqalign{
&b(\sigma =0)=b(\sigma =\sigma _{max})=0,\cr
&\phi ^\prime (\sigma =0)=\phi ^\prime (\sigma =\sigma _{max})=0.\cr
}\eqn\bncc$$
(The latter two conditions
follow from the continuity of the solution - the parameter
$\sigma$ is related to the time $t$ inside the bubble by $\sigma=-it$,
so we have at $sigma =t=0$ that $(\partial \phi /\partial \sigma)
= i (\partial \phi / \partial t)$, with $\phi(t)$ and $\phi(\sigma)$
real.)

In the flat space case (where the coupling to gravity is turned off),
$\sigma _{max}=+\infty $ and the bounce action starts in the false
vacuum, with $\phi (\sigma =+\infty )=\phi _{fv}.$ Because of the
first derivative term (which behaves as a sort of friction term),
$V[\phi (\sigma =0)]$ is slightly higher than $V[\phi _{fv}].$
With gravity turned on $\sigma _{max}$ is finite, and
the Euclidean instanton does not quite start in the false vacuum
but rather partially into the barrier. As the coupling to gravity
is further increased, the Coleman--de Luccia instanton approaches
the Hawking-Moss instanton---in other words, $\phi (\sigma =0)$ and
$\phi (\sigma =\sigma _{max})$ approach the maximum of the potential
\refmark{\jensen} . For a discussion of the interpretation
of the Hawking-Moss instanton, we refer the reader to
refs. \lhm.

If one continues the Coleman--de Luccia bounce to a spacetime with a
Lorentzian spacetime signature, the functions $\phi (\sigma )$ and
$b(\sigma )$ describe the solution in region II. In region I (inside
the bubble), the field $\phi $ rolls toward the true vacuum. In
region IV the field rolls toward the false vacuum, undergoing
decay and an infinite number of oscillations. [In the maximal
Lorentzian extension, regions III and V mirror the behavior in
regions I and IV, but these regions lie in the domain of
Euclidean evolution.]

We shall be primarily concerned with the case where the size
of the bubble is small so that a good deal of the Euclidean
bounce is very nearly de Sitter space. In this case we may
use the Bunch-Davies vacuum (for a scalar field of constant
mass in de Sitter space) as an initial condition for
the fluctuations in the scalar field at $t =0.$

There are two
unsettling aspects of the Coleman-de Luccia instanton.
The first
is the global character of the tunneling process. All of
de Sitter space is involved. There is no asymptotic region
in which the deviations in the fields from their false vacuum
values vanishes. Formally, this is the result of constructing
the instanton from a closed geometry (i.e., choosing a time
variable for analytic continuation such that the constant time
hypersurfaces are compact). One would prefer an instanton
based on the `flat' coordinatization of de Sitter space,
in which the tunneling event would occur within a Hubble radius,
and in which the instanton would have a tail
falling off rapidly
outside the Hubble radius. Such an
instanton would seem more physical, but unfortunately no
such instanton is known.
The second unsettling aspect is that these instantons involve
continuing the spacetime metric as well as the scalar field.
In a sense one is considering complex metric configurations.
So there is implicitly some sort of integration of
spacetime metrics being performed (i.e., a path integral for
quantum gravity), which one is then distorting to run
over a complex
saddle point. We shall in this paper adopt a conservative
view, namely
imagine that we are working in the regime where
the nucleated bubble is
much smaller than the Hubble radius $H^{-1}$, so that the
gravitational effects are small. Our
bubble solutions are then well described by the Minkowski space
solutions, and so (hopefully) little affected by these
quantum gravitational issues.

\chapter{A New `Thin Wall' Limit}

This open inflation scenario can be most simply explored in the following
regime:

(1.) The bubble nucleation rate is small, so that it is reasonable to
suppose that the universe became very accurately
described by de Sitter space,
and the quantum field mode functions became well approximated by
the Bunch-Davies vacuum
modes.

(2.) The size of the nucleated bubble is small
compared to $H^{-1}$. This allows us to treat the nucleating bubble
as a geometrical `point' on the scales $\sim H^{-1}$
of primary interest, and to suppose that it has very little effect on
smaller scales.

(3.) The bubble wall is thin. This allows us to match modes
across a geometrical null surface, rather than evolve scalar field
modes through the complicated structure of a bubble wall.

(4.) The potential $V(\phi)$ around
the region $\phi_n$ is accurately approximated
as linear. This again affords a useful technical simplification.
Since we wish `old' inflation to be followed
by `new' or `slow roll' inflation it seems reasonable  to
assume that
$V(\phi)$ is  a slowly varying (i.e., approximately linear)
function around $\phi_n$.

(5.) The vacuum energy in the
false vacuum $F$ is nearly equal to that along the light cone
$n$. This requires that
$\Delta V \ll V$ (see Figure 3). Again this allows us to
treat the background de Sitter geometry as fixed in the
matching of modes across the bubble wall.

It is not obvious that such a regime exists, and
it is the purpose of this section to show that it does.
The `thin wall' solution that we are looking for is not the standard one,
described for example by Coleman and De Luccia,
which holds when the false vacuum
$F$ is nearly degenerate with the true vacuum. Clearly
we do not want to be in this regime, since then we would
not have much `new' inflation inside the bubble.
In this section we find an
analytic solution for the Euclidean `bounce' and
classical Minkowski bubble which holds in the
regime we want.

We consider the case where the mass $m$ of the field $\phi$
in the false vacuum $F$ is large
so  the potential well
around $\phi_i$ becomes very narrow, but its height
$\Delta V$ is fixed (see Figure 3).
Our solution is exact in the limit as $m$ becomes infinite.
This is not in itself a case of much physical interest, but
we do expect the infinite $m$ solution
to be a reasonable approximation to the case
we {\it are} most interested in, where $m$ is substantially larger
than $H$.

Anticipating that the final bubble size will be much smaller
than $H^{-1}$, we use eqn. \bncb , in the limit $H\sigma \ll 1$.
The equation for $b(\sigma)$ tells us that at small $\sigma$,
$b \sim \sigma \sim r$, the usual radial variable in flat space.
The scalar field equation then reduces to  the
flat spacetime equation for the Euclidean bounce:
$$
\phi'' + {3\over r} \phi' = - V_{,\phi} = -\mu^3.
\eqn\neila
$$
Here we have assumed the potential is linear over
the entire relevant range. This is true right up to the
beginning of the `well' around $\phi_i$, which we assume to
be very narrow.
The bounce starts with $\phi=\phi_n$ at $r=0$. $\phi$
evolves according to eqn. \neila ~ up to the edge of the `well',
and we have
$$
\phi = \phi_n - {\mu^3 \over 8} r^2.
\eqn\neilb
$$
As usual, the solution $\phi(r)$ may be thought of as the trajectory
of a particle moving in a potential $-V(\phi)$ with $r$ playing the role
of time. With this interpretation, the solution of the equation
is then clear. $\phi$ accelerates according to \neilb, and hits
the barrier $-\Delta V$ at some value of $r$ we shall call $r_b$,
its kinetic energy being
converted into potential energy. (When the potential well is very narrow,
$\phi$ is decelerated to a standstill in a very small interval of
$r$, during which the loss of
energy from the damping term is insignificant).
Thus one finds using eqn. \neilb ~ and setting $\phi =\phi_i$ that
at $r_b$,
$$
{1\over 2} \phi'^2 = (\phi_n-\phi_i){ \mu^3 \over 4} = \Delta V.
\eqn\neilc
$$
This equation determines $\phi_n$, and then eqn. \neilb ~
determines $r_b$ so that
$$
\phi_n= \phi_i + 4 {\Delta V \over \mu^3} ,\qquad \qquad r_b=
\sqrt{32 \Delta V \over \mu^6}.
\eqn\neild
$$
The Euclidean action for this solution is:
$$
S_E = 2 \pi^2 \int_0^{r_b} r^3 dr \bigl[{1\over 2} \phi'^2 +
(V(\phi)-V(\phi_i))\bigr]
= {512 \over 3} \pi^2 {\Delta V^3 \over \mu^{12}}.
\eqn\neile
$$

Let us check the consistency of this solution.
The semiclassical
approximation
is only likely to be reliable in the case where the Euclidean
action is large, which is true if we take  $\Delta V \sim \mu^4$.
The mass $m$ has to be large in order that we can treat
the false vacuum potential well as narrow---approximating
it as $(const.)+m^2 (\phi-\phi_r)^2$, we require that
the width of the well $|\phi-\phi_r| \sim \sqrt{\Delta V}/m$
be much less than  $\phi_n = 4 \Delta V \mu^{-3}$. Taking
$\Delta V \sim \mu^4$, we see the narrow well approximation
should hold well if $m \gg \mu$.
Next we compare  the bubble size
$r_b \sim \mu^{-1}$ (Eqn. \neild )
to the Hubble radius $H^{-1} \sim 3 m_{pl} V^{-{1\over 2}}
\sim m_{pl}^{1\over 2} \mu^{- {3\over 2}} \gg  \mu^{-1}$ if
$\mu \ll  m_{pl}$. This justifies our use of the flat
spacetime approximation in eqn. \neila.

These results mean that conditions 1 and 2 above are
satisfied. Condition 3 follows after noting that the
classical  solution for the expanding bubble is given
by exactly the same  solution (analytically continued)
as the Euclidean bounce
outside the forward light cone of the nucleated bubble's
center, so conditions 3 and 2 are equivalent. Condition
4 is satisfied by construction, and condition 5 holds
as long as $\Delta V \ll  V$, which is also true
if
$\mu \ll m_{pl}$ (recall from section 2
that the total range of $\phi$
from $\phi_i$ to $\phi_r$ has to be of order $m_{pl}$
to get a reasonable value for $\Omega_0$).

We have shown that there exists a regime in which the
nucleating bubble, and the
bubble wall become `geometrical,' and the problem simplifies.
This is the case if the mass in the false vacuum $m$ obeys $m^2 \gg H^2$.
However this case appears to be technically more complicated than the
case where $m^2 = 2H^2$, for which a minimally coupled
scalar field has the same dynamics as a conformally coupled, massless
field. So for the remainder of the paper we shall restrict
ourselves to $m^2=2H^2$, but nevertheless treat the bubble wall
as `thin'. We do this simply in order to be able to pursue the
calculation right through to the end in a simple way.
We do not anticipate any difficulties of principle
in extending the treatment to $m^2 \gg H^2$ (see note added).

The $m^2= 2H^2$ case may actually be a
reasonable approximation
in the case of a two-field inflation theory, in which the
first field, undergoes a bubble nucleation
transition, in which it has a large mass squared
in both phases, and there is no `slow roll' phase.
If this couples to a second field, in such a way that the
latter has
$m^2=2H^2$ in the false vacuum,
and $m^2 \approx 0$ after the tunneling event, then
the second field  can subsequently undergo a
`slow roll' transition. As far as the second field is concerned,
the bubble wall may  then be treated as `thin'.
However we do not wish to enter into the complexities of
two-field inflation here, and in fact
regard it as a more interesting problem
to extend the discussion to a single field,
with $m^2 \gg H^2$ (see note added).

\chapter{$SO(3,1)$ Invariant Coordinates}

In this section we set up the coordinate systems and
mode expansions most convenient for describing the
expanding bubble and the perturbations about the perfect
expanding bubble solution. The coordinate system that maximally
exploits the $SO(3,1)$ invariance divides the $(3+1)$--dimensional
spacetime into three regions, indicated in Figure 2, in much the
same way as Rindler coordinates divide $(1+1)$--dimensional
Minkowski space into four regions. Region I consists of the
interior of the forward light cone of the origin $O,$ region
II consists of points with a spacelike separation relative
to $O,$ and region III consists of the interior of the backward
light cone of $O.$

In many parts of our calculations, we shall consider only the
$s$-wave sector. Because of the high underlying symmetry of
the perturbations produced in the expanding bubble background,
all of the density perturbations may be deduced from the
$s-$wave sector. We calculate the two-point function choosing
the origin as one of the two points, in which case only the
$s-$wave sector contributes. Because of homogeneity and isotropy,
the two-point function for all positions of the two points is
determined, and thus the behavior of all other modes with higher
angular momenta.

For region I, the metric has the line element
$$ ds^2=-dt^2+a^2(t)~\bigl[ d\xi ^2+\sinh ^2\xi ~
(d\theta ^2+\sin ^2\theta ~d\phi ^2)~\bigr] ,
\eqn\leone$$
and $\sqrt{-g}=a^3(t)~\sinh ^2\xi \sin \theta .$
[For the special cases of de Sitter space and Minkowski
space, $a(t)=H^{-1}\sinh [Ht]$ and $a(t)=t,$ respectively.]
In region I, the equation of motion for the scalar field of
the unperturbed bubble solution is
$$ \ddot \phi _0(t)+3 {\dot a(t)\over a(t)}~\dot \phi _0(t)
+V'[\phi _0(t)]=0\eqn\vvai$$
and the Einstein equation is
$$ {\dot a^2(t)\over a^2(t)} =
{1\over a^2(t)}+{8\pi \over 3m_{pl}^2}
\left\{ {1\over 2}
\left( {\partial \phi _0\over \partial t}\right) ^2
+V[\phi _0(t)]\right\}.\eqn\vvbi $$

For region II, the metric has the line element
$$
ds^2= d\sigma ^2+b^2(\sigma )~\bigl[ -d\tau ^2+\cosh^2\tau ~
(d\theta ^2+\sin ^2\theta ~d\phi ^2)~\bigr] ,
\eqn\letwo
$$
and $\sqrt{-g}=b^3(\sigma )\cosh ^2\tau \sin \theta .$
[For the special cases of de Sitter space and Minkowski
space, $b(\sigma )=H^{-1}\sin [H\sigma ]$ and $b(\sigma )=\sigma ,$
respectively.]
Likewise, in region II the equation of motion for the unperturbed
bubble solution is (denoting $\partial_\sigma \phi = \phi^{\prime}$)
$$ \phi _0^{\prime \prime }(\sigma )+{3b'(\sigma )\over
b(\sigma )}~\phi _0^\prime (\sigma )
-{\partial V\over \partial \phi} [\phi _0(\sigma)]=0,\eqn\vvaii$$
and the Einstein equation is
$$ {b^{\prime ~2}(\sigma )\over b^2(\sigma )}=
{1\over b^{~2}(\sigma )}+{8\pi \over 3m_{pl}^2} \left\{
{1\over 2}\left( {\partial \phi _0\over \partial \sigma }\right) ^2
-V[\phi _0(\sigma )]\right\} =0.\eqn\vvbii $$
Note that these are exactly the same equations as those
for the Euclidean `bounce'.
Also, eqns. \vvaii ~ and \vvbii ~  are related to
eqns. \vvai ~ and \vvbi ~  by the substitutions $\xi=\tau+ i \pi/2$,
$t=i\sigma$, $a=ib$.

The functions
$\phi _0(t)$ and $\phi _0(\sigma )$
form the `perfect' (i.e, $SO(3,1)$ invariant) bubble
solution in regions I and II, respectively. If they are
to  match smoothly
on the future  light cone of the nucleated bubble ($\sigma=t=0$),
 we must require that
$$\eqalign{ \phi _0(\sigma =0)&=\phi _0(t=0),\cr
\dot \phi _0(t=0)&=\phi '_0(\sigma =0).\cr }\eqn\vvva$$
We now consider, to linear order in $\tilde \phi ,$
small perturbations of the perfect bubble
$$ \phi =\phi _0+\tilde \phi .\eqn\vvc$$

For region I, the d'Alembertian takes the form
$$\eqalign{
\Box =& {1\over \sqrt{-g}}\partial _\mu [\sqrt{-g}~g^{\mu \nu }
\partial _\nu ]\cr
=&\partial _t^2+{3\dot a(t)\over a(t)}~\partial _t\cr
& ~~~
-{1\over a^2(t)}\left[
\partial ^2_\xi +2\coth  \xi ~\partial _\xi
+{1\over \sinh ^2\xi }\left( \partial _\theta ^2
+\cot \theta ~\partial _\theta +{1\over \sin ^2\theta }\partial ^2_\phi
\right) \right]\cr
=&\partial _t^2+{3\dot a(t)\over a(t)}~\partial _t
-{1\over a^2(t)}\left[
\partial ^2_\xi +2\coth  \xi ~\partial _\xi
+{1\over \sinh ^2\xi }\left(
-{\bf L}^2
\right) \right].\cr
}\eqn\daone$$

Using separation of variables, we find the
mode expansion for the solutions of the
wave equation
$$[~ \Box +m^2(t)]\tilde \phi =0\eqn\vve$$
where $m^2(t)=V^{\prime \prime }[\phi _0(t)].$
Writing
$$\tilde \phi (t,\xi ,\theta ,\phi )
=T(t;k_{h})~R(\xi ;k_{h},l)~Y_{lm}(\theta ,\phi ),\eqn\exa$$
one obtains
$$\eqalign{
&\left[ \partial _t^2+{3\dot a(t)\over a(t)}\partial _t +{1\over a^2(t)}
k^2_{h}+m^2(t)
\right]T(t;k_{h})=0,\cr
&\left[ \partial _\xi ^2 +2\coth \xi ~\partial _\xi
-{l(l+1)\over \sinh ^2\xi }+k_{h}^2\right] R(\xi ;k_{h},l)=0.\cr
}\eqn\lista$$
For region II one has
$$
\Box =-\partial _\sigma ^2 -{3b'(\sigma )\over b(\sigma )}\partial _\sigma
+{1\over b^2(\sigma )}\left[ \partial _\tau ^2 +2\tanh \tau ~
\partial _\tau -{1\over \cosh^2 \tau }(-{\bf L}^2)\right] .
\eqn\www$$
Writing
$$
\tilde \phi (\sigma ,\tau , \theta, \phi )=
S(\sigma ;\omega _{dS})~Q(\tau ;\omega _{dS},l)Y_{lm}(\theta ,\phi ),
\eqn\wwa$$
one obtains the equations
$$
\eqalign{
&S^{\prime \prime }+{3b^\prime (\sigma)\over b(\sigma )}S'
+\left[ {\omega _{dS}^2\over b^2(\sigma )}-m(\sigma )^2
\right]S=0,\cr
&\ddot Q+2\tanh \tau ~\dot Q +\left[
{l(l+1)\over \cosh ^2\tau }+\omega ^2_{dS}\right] Q=0.\cr }
\eqn\wwb$$

Define
$$\tilde \phi (t, \xi ;l,m)=\int _{S^2} d\Omega ~Y_{lm}^*(\Omega )
\tilde \phi (t,\xi ,\theta ,\phi ),
\eqn\wwaa$$
and define $\tilde \phi (\sigma , \tau ; l, m)$ similarly.

In region I one may express the most general solution
in the form
$$\eqalign{
\tilde \phi (t, &\xi ; l, m )=
\int _0^\infty d\zh ~R(\xi ;\zh,l)\cr
&\times \left[
A^{(+)}_I(\zh,l,m) ~T^{(+)}(t;\zh) +
A^{(-)}_I(\zh,l,m) ~T^{(-)}(t;\zh)
\right] .\cr }
\eqn\xxa$$
We define $\kh ^2=\zh ^2+1$ for
future convenience, and $(+)$ and $(-)$ label two
linearly independent solutions for the temporal evolution.
Likewise, in region II we may expand the most general solution
in the form
$$\eqalign{
\tilde \phi (\sigma ,\tau ;l,m)=&
\int _{-\infty }^{+\infty }d\zeta _{dS}~
S(\sigma ;\zd )\cr
&\times \left[
A^{(s)}_{II}(\zd ;l,m)~Q^{(s)}(\tau ;\zd ,l)
+
A^{(a)}_{II}(\zd ;l,m)~Q^{(a)}(\tau ;\zd ,l)\right]
,\cr } \eqn\xxb$$
where $\omega _{dS}^2= \zeta _{dS}^2+1.$

We now proceed to calculate the properties of the mode
functions. Note that the hyperbolic spherical functions
$R_l(\xi ;\zh )$ and the de Sitter functions
$Q^{(s)}(\tau ;\zd ,l)$ and $Q^{(a)}(\tau ;\zd ,l)$
are universal, because they are completely determined by
the $SO(3,1)$ symmetry; these functions do not depend on
$a(t),$ $b(\sigma ),$ $m^2(t),$ and
$m^2(\sigma ).$ By contrast, the coefficient functions
$T_l(t;\zh )$ and $S_l(\sigma ;\zd )$ are not universal and
do depend on the functions
$a(t),$ $b(\sigma ),$ $m^2(t),$ and $m^2(\sigma ).$

{\bf Hyperbolic Spherical Functions.} A nice discussion of
the hyperbolic spherical functions with references to
earlier work appears in ref. \bander .
These functions satisfy the equation
$$\left[
{d ^2\over d\xi ^2}+2\coth \xi ~ {~d \over d \xi }+
(\zh ^2 +1)-{l(l+1)\over \sinh ^2\xi }
\right] R_l(\xi ;\zh )=0\eqn\hsa$$
where $\kh ^2=\zh ^2+1.$ It is convenient to rewrite \hsa ~ as
$$\eqalign{
\Biggl[
\sinh ^2\xi ~{d^2\over d\cosh \xi ^2}&+3\cosh \xi ~
{d \over d \cosh \xi }
+(\zh ^2 +1)-{l(l+1)\over \sinh ^2\xi }
\Biggr] \cr &\times R_l(\xi ;\zh )=0.\cr }\eqn\hsb$$
It may be shown \refmark{\dolginov } that the functions
$$ R_l(\xi ;\zh )=N_l(\zh )\cdot (-)^{l+1}\cdot \sinh ^l \xi ~
{d^{l+1}\over d\cosh \xi ^{l+1}} \cos (\zh \xi )\eqn\hsc $$
satisfy \hsb , where
$$ N_l(\zh )={1\over
\sqrt{ {\pi \over 2} \zh ^2 (\zh ^2+1^2) (\zh ^2+2^2) \ldots
(\zh ^2+l^2) } }\eqn\hsd$$
and the orthogonality relation
$$
\int _0^\infty d\xi ~\sinh ^2\xi ~
R_l(\xi ;\zeta _1 )~R_l(\xi ;\zeta _2)=\delta (\zeta _1-\zeta _2)\eqn\hsf
$$
and the completeness relation
$$
\int _0^\infty d\zh ~R_l(\xi _a;\zh )~R_l(\xi _b;\zh )=
{\delta (\xi _a-\xi _b)\over \sinh ^2 \xi _a}
\eqn\hsg $$
hold. Note that
$$
R_0(\xi ,\zh )=\sqrt{2\over \pi }\cdot{\sin [\zh \xi ]\over
\sinh \xi }.\eqn\hsgga$$

{\bf de Sitter Functions.} The de Sitter functions satisfy
the equation
$$\left[
{d ^2\over d\tau ^2}+2\tanh \tau ~ {~d \over d \tau }+
(\zd ^2 +1)+{l(l+1)\over \cosh ^2\tau }
\right] Q_l(\tau ;\zd )=0,\eqn\dsa$$
which can be rewritten as
$$\left[
\cosh ^2\tau ~{d ^2\over d\sinh \tau ^2}+3\sinh \tau ~ {~d \over d \sinh \tau }
+ (\zd ^2 +1)+{l(l+1)\over \cosh ^2\tau }
\right] Q_l(\tau ;\zd )=0\eqn\dsb$$
with $\omega _{dS}^2=\zd ^2+1.$
We demonstrate that
$$Q_l^{(\pm )}(\tau ;\zd )=n_l(\zd )~\cosh ^l\tau ~
{d^{l+1}\over d\sinh \tau ^{l+1}} ~ e^{\pm i\zd \tau }.
\eqn\dsc$$
satisfies \dsa .\footnote\dagger{
It is readily verified that
$Q_0^{(\pm )}(\tau ;\zd )=e^{\pm i\zd \tau }/\cosh [\tau ]$
satisfies \dsa  ~ with $l=0.$
We generalize this result inductively to all $l.$ For this it is
convenient to rescale $z=\sinh \tau $ and $Q_l(\tau )=\cosh ^l\tau ~
w_l(z),$
so that \dsa ~  becomes
$$(z^2+1)w_l^{\prime \prime }(z)+(2l+3)zw'_l(z)+
[(\zd ^2+1)+l(l+2)]w_l(z)=0.\eqn\dsf$$
If $w_l$ satisfies \dsf ~ for $l,$
then $w^\prime _l$ satisfies \dsf ~ for $(l+1).$}

{\bf Peculiar Mode Functions.} We now discuss
the other mode functions $T_l(t;\zh )$ and
$S_l(\sigma ;\zd ),$ which depend on the particular
choices for $a(t),$ $b(\sigma ),$ $m^2(t),$ and
$m^2(\sigma ).$ In the general case, for the
expanding bubble, the behavior of these functions
depends on the particular choice of potential and
will have to be evaluated numerically. However,
for certain special cases, we may solve for these
functions analytically---in particular for
a field of constant mass (i.e., whose mass does
not vary with spacetime position) in de Sitter space
and in Minkowski space.

There are two reasons for considering these special
cases. First of all, by considering a massless field
in Minkowski space, we may determine matching conditions
at the light cone that are generally valid. More
specifically, we would like to determine the matrix
elements of the integral transform
$$\eqalign{&%
\pmatrix{ A^{(+)}_I(\zh ; l, m )\cr
A^{(-)}_I(\zh ; l, m )\cr }=
\int _{-\infty }^{+\infty }d\zd ~\cr
&~~~~ \times \pmatrix{
{\cal T}^{(+,s)}_l(\zh \vert \zd )
&{\cal T}_l^{(+,a)}(\zh \vert \zd )\cr
{\cal T}^{(-,s)}_l(\zh \vert \zd )
&{\cal T}_l^{(-,a)}(\zh \vert \zd )\cr }
\times
\pmatrix{
A^{(s)}_{II}(\zd ; l, m )\cr
A^{(a)}_{II}(\zd ; l, m )\cr }\cr }\eqn\nna$$
relating the expansion coefficients in regions I and II.

This task would be quite simple if the individual mode
functions in regions I and II did not become singular
as $t\to 0$ and $\sigma \to 0,$ respectively. One would
simply require that $\phi $ and its transverse derivative
be continuous across the light cone. However, each of the
individual modes becomes highly oscillatory as one approaches
the light cone. Therefore, wave packets that are well behaved
near the light cone must be constructed as suitable smeared
superpositions of individual modes.

For matching the two expansions across the light
cone, only the asymptotic behavior of
$T_l(t;\zh )$ and $S_l(\sigma ;\zd )$ for small $t$
and $\sigma $ is relevant. Therefore, we may
calculate the matching for Minkowski space,
with $a(t)=t$ and $b(\sigma )=\sigma ,$
and the result obtained will be generally
applicable, for all choices for the functions
$a(t),$ $b(\sigma ),$ $m^2(t),$ and $m^2(\sigma ),$
provided that the mode functions
$T_l(t;\zh )$ and $S_l(\sigma ;\zd )$
are normalized to have the same asymptotic behavior
near the light cone as the functions to calculate the
matching in Minkowski space.

{\bf Mode functions for Minkowski space.}
In Minkowski space, eqn. \lista ~  becomes
$$ \ddot T+{3\over t}~\dot T+
\left[ m^2 +{\kh ^2\over t^2}\right] T=0, \eqn\wwk$$
which with the change of dependent variable
$T(t)=Y(t)/t$  becomes
$$ \ddot Y+{1\over t}~\dot Y+
\left[ m^2+{\zd ^2\over t^2}\right] Y=0 \eqn\wwl$$
where $\kh ^2=\zh ^2+1.$ The solutions are Bessel functions of
imaginary order $J_{\pm i\zh }(mt),$ and
$$T_l^{(\pm)}(t;\zh )={1\over t}~J_{\pm i\zh }(mt).\eqn\wwm$$

For the massless case $m^2=0,$ one has
$$ T_l^{(\pm )}(t;\zh )=t^{\pm i\zh -1}=
{1\over t}e^{\pm i\zh ~\ln t}.\eqn\eea $$
Similarly in region II, for Minkowski space \wwb b becomes
$$S^{\prime \prime }+{3\over \sigma }S'+\left[
{\omega _{dS}^2\over \sigma ^2}-m^2\right] S=0.
\eqn\wwn $$
Making the substitution $S(\sigma )={Z(\sigma )/\sigma },$ one obtains
$$ Z^{\prime \prime }+{1\over \sigma }Z'+
\left[ {\zd ^2\over \sigma ^2}-m^2
\right]Z=0 \eqn\wwo$$
where $\omega _{dS}^2=\zd ^2+1.$ The solutions are
modified Bessel functions of imaginary order, and
$$S(\sigma )={1\over \sigma }K_{\pm i\zd }(m\sigma ).  \eqn\wwp$$
For small $\sigma$
$$
S(\sigma )\approx \sigma ^{\pm i\zd -1}
={e^{\pm i\zd {\rm ln}[\sigma ]}
\over \sigma }.\eqn\eec $$

For $m^2>0$ only
one solution has acceptable behavior for large $\sigma .$
For the massless case, there are twice as many acceptable solutions,
since both solutions oscillate as $\sigma \to \infty .$
This doubling has a simple explanation.
For the massive case, the light cone regarded as a Cauchy
surface is equivalent to the Cauchy surface defined by
$\tau =0,$ because for a massive field, at least
massive in the limit $\sigma \to \infty ,$ all of the
information on the $\tau =0$ surface must eventually
propagate into the light cone. For a massless field
this is not so. There is a set of modes that
propagate to null infinity ahead of the light cone, and
hence never enter the light cone.

{\bf Coordinate Systems for de Sitter Space.} Before giving
the mode functions for de Sitter space, to establish
notation we write down the four sets of coordinates
for de Sitter space used in the course of our calculations---%
which we shall call `hyperbolic,' `flat,' `closed,' and
`embedded' coordinates---and the transformations between them.

One may construct $(3+1)$--dimensional de Sitter space as an
embedding in $(4+1)$--dimensional flat Minkowski space, defined
by the equation
$$\bar x^2+\bar y^2+\bar z^2 +\bar u^2-\bar w^2=1.\eqn\cca$$
We call the coordinates $\bar w,$ $\bar u,$ $\bar x,$ $\bar y,$
$\bar z$ `embedded' coordinates.

The `closed' coordinates  are then given by
$$\eqalign{
\bar w&=\sinh [t_{cl}],\cr
\bar u&=\cosh [t_{cl}]~\cos [\chi _{cl}],\cr
\bar z&=\cosh [t_{cl}]~\sin [\chi _{cl}]\cos [\theta ],\cr
\bar x&=\cosh [t_{cl}]~\sin [\chi _{cl}]
\sin [\theta ]\cos [\phi ],\cr
\bar y&=\cosh [t_{cl}]~\sin [\chi _{cl}]
\sin [\theta ]\sin [\phi ].\cr }\eqn\cca$$
They cover all of de Sitter space,
and the line element is
$$ds^2=-dt_{cl}^2+\cosh ^2[t_{cl}]\cdot (d\chi _{cl}^2
+\sin ^2 \chi _{cl} d\Omega ^2_{(2)}).
\eqn\ccaa$$

In the `flat' coordinates, the constant time hypersurfaces
are null planes in the five-dimensional
Minkowski space, into which the de Sitter space is
embedded. More specifically,
$$\eqalign{t_f &={\rm ln}[ \bar w+\bar u ],\cr
r_f &={\bar r\over \bar w+\bar u},\cr }\eqn\cce$$
and the line element is
$$ds^2=-dt_f^2+e^{2t_f}\cdot
\bigl[ dr_f^2+r_f^2d\Omega ^2_{(2)}\bigr] .
\eqn\cccea$$
where $\bar r= (\bar x^2+\bar y^2+\bar z^2)^{1/2}.$
The flat coordinates cover only the half of de
Sitter space defined by $\bar w+\bar u>0.$ The other
half is covered by another set of `flat' coordinates.

Next we discuss the
region II `hyperbolic' coordinates
$$\eqalign{ \bar w&=\sinh [\tau ] ~\sin[\sigma],\cr
\bar u&=\hskip 0.6in \cos [\sigma ],\cr
\bar x&=\cosh [\tau ]~\sin [\sigma ]~\cos [\theta ],\cr
\bar y&=\cosh [\tau ]~\sin [\sigma ]~\sin [\theta ]~
\cos [\phi ],\cr
\bar z&=\cosh [\tau ]~\sin [\sigma ]~\sin [\theta ]~
\sin [\phi ],\cr
}\eqn\ccg$$
in which the line element is
$$ds^2=d\sigma ^2+\sin ^2[\sigma ]\cdot
\bigl[ -d\tau ^2+\cosh ^2[\tau ]d\Omega ^2_{(2)}
\bigr] .\eqn\ccgzz$$

Finally we have the region I `hyperbolic' coordinates
$$\eqalign{
\bar w&= \sinh [t]\cosh [\xi ],\cr
\bar u&= \cosh [t],\cr
\bar x&= \sinh [t]\sinh [\xi ]\sin [\theta ]\cos [\phi ],\cr
\bar y&= \sinh [t]\sinh [\xi ]\sin [\theta ]\sin [\phi ],\cr
\bar z&= \sinh [t]\sinh [\xi ]\cos [\theta ],\cr
}\eqn\ccgga$$
in which the line element is
$$
ds^2=-dt^2+\sinh ^2[t]\cdot
\bigl[ d\xi ^2+\sinh ^2[\xi ]d\Omega ^2_{(2)}\bigr]
\eqn\ccggab$$
These coordinates cover the forward light cone of the point $P$
described by the embedded coordinates $\bar u=1,$
$\bar x=\bar y=\bar z=\bar w=0.$ To cover all of
de Sitter space with hyperbolic coordinates, three
more coordinates patches of a form similar to region I
are required. Region III consists of the backward
light cone of $P.$ Region IV consists of the forward
light cone of the antipodal point of $P,$ which we call
$\bar P,$ and which is described by the embedded
coordinates $\bar u=-1,$ $\bar x=\bar y=\bar z%
=\bar w=0.$ Finally, region V is the backward light
cone of $\bar P.$

{\bf Peculiar Functions for de Sitter Space.}
For de Sitter space in region II, the scale factor
takes the form $b(\sigma )=H^{-1}\sin [H\sigma ].$
For simplicity we set $H=1.$ Eqn. \wwb a becomes
$$S^{\prime \prime }(\sigma )
+3\cot [\sigma ]~S^{\prime }(\sigma )-m^2S
+{[\zd ^2+1]\over \sin ^2[\sigma ]}~S(\sigma )=0,
\eqn\yya$$
where $0\le \sigma \le \pi .$ Let $x=\cos \sigma $ and
$S={F/ \sin [\sigma ]}={F/(1-x^2)^{1/2}},$ so that eqn.
\yya ~ becomes
$$ {d\over dx}\left[ (1-x^2)
{dF\over dx}\right]
+\left[ (2-m^2)+{\zd ^2\over 1-x^2}
\right]F=0,\eqn\yyb$$
which is the Legendre differential equation, with
linearly independent solutions $P_\nu ^{\pm i\zd}(x)$
and $Q_\nu ^{\pm i\zd}(x)$ with $\nu (\nu +1)=(2-m^2).$

For $m^2=2,$ which we shall call the `conformal mass' case
(because the quadratic term in the action has the same
effect as no mass with conformal coupling), we solve
eqn. \yyb ~ by making the change of variable $x=\tanh [u],$ with
$-\infty <u<+\infty ,$ so that eqn. \yyb ~ becomes
$${d^2F\over du^2}+
(2-m^2)\sech ^2[u] ~F
+\zd ^2~F=0,\eqn\yyc$$
which for $m^2=2$ has the solutions
$F=e^{\pm i\zd u}.$ For the massless case $(m^2=0)$ one has
$F=(i\zeta -\tanh [u])e^{\pm i\zeta u}.$

\chapter{Initial Conditions}

One of the remarkable properties of a de Sitter space
background is its ability to erase intial conditions
through exponential expansion. After a sufficient number
of e-foldings, the state of the scalar field within a
Hubble volume becomes almost completely determined
and almost completely independent of the intial
conditions.

In this section we discuss the quantum fluctuations
of the inflaton field prior to bubble nucleation, when
the inflaton field is stuck in the false vacuum.
There exist extensive discussions in the literature
of the quantum fluctuations in a
 free scalar field of fixed
mass in de Sitter space. The principal result of
this section is an expansion of the
two-point function describing the Bunch-Davies vacuum
(see ref. \birrell ~ and references therein)
in terms of the region II hyperbolic mode functions.
To follow the evolution of the quantum fluctuations
through the bubble wall, it is necessary to describe
the initial state in this manner.

The choice of the Bunch-Davies vacuum as a reasonable
initial condition can be justified in the following
way. In de Sitter space, if one follows any scalar field mode
in the flat spatial slicing,
its proper wavelength
 begins inside the Hubble radius, and is exponentially stretched
outside the Hubble radius as time proceeds. So the early evolution
is effectively the same as in Minkowski spacetime.
The energy density in the scalar field fluctuations is
the renormalized sum of the energy in each field mode - if this
is to be finite, it is clear that the very short wavelength modes
must begin in their ground state (i.e. the Minkowski vacuum).
It follows that after a sufficient number
of expansion times, the state of the scalar field (as
observed over a fixed physical volume) becomes independent
of the initial state of the scalar field, because all one
sees are the modes which began in their ground state.
This `evolved Minkowski space vacuum' is the
Bunch-Davies vacuum.

For the two-point function in the Bunch-Davies vacuum
$\vert 0_{BD}\rangle $ we shall use the Wightman function
$$G^{(+)}(X,X^\prime )
= \langle 0_{BD}\vert
\hat \phi (X)\hat \phi (X^\prime )\vert 0_{BD}\rangle .\eqn\jja$$
where $X=(\bar w, \bar u, \bar x, \bar y, \bar z)$ are the
`embedded' coordinates of the previous section.
Since $G^{(+)}$ is
invariant under the action of the connected part
of $SO(4,1),$ we express $G^{(+)}$ in terms of
the invariant
$$I(X,X^\prime )=
-\bar w\bar w^\prime
+\bar u\bar u^\prime
+\bar x\bar x^\prime
+\bar y\bar y^\prime
+\bar z\bar z^\prime -i\epsilon ~\varepsilon(X,X^\prime )
\eqn\jjb$$
Note that $I(X,X^\prime )>+1$ when $X$ lies inside
either the forward or backward light cone of $X^\prime.$
$I(X,X^\prime )=+1$ when $X$ lies on the light cone of
$X'.$ When
$-1<I(X,X^\prime )<+1,$ there exists a spacelike geodesic
connecting $X$ and $X^\prime .$ When $I(X,X^\prime )=-1,$
then $X$
lies on the light cone of the antipodal point of
$X^\prime .$ Finally, when $I(X,X^\prime )<-1,$ $X$ lies
in the interior of the light cone of the antipodal point
of $X^\prime ,$ and
in this case there is no geodesic connecting $X$ to $X^\prime .$
We add a small imaginary part to $I(X,X^\prime )$ to indicate the
time ordering of $X$ and $X^\prime .$
$\varepsilon (X,X^\prime )$ vanishes when $I(X,X^\prime )<1,$
is equal to $+1$ when $X$ lies inside the forward light cone
of $X^\prime ,$ and is equal to $-1$ when $X$ lies inside
the backward light cone of $X^\prime .$

We have
$$G^{(+)}(X,X^\prime )={({1\over 4}-\nu^2)
\over 16\pi }\cdot \sec [\pi \nu ]\cdot
F\left[ {3\over 2}-\nu ,{3\over 2}+\nu ;
2; {I(X,X^\prime )+1\over 2}\right]
\eqn\jjbb$$
where $\nu ^2={9\over 4}-m^2.$ Except for special degenerate
values of $\nu ,$ the hypergeometric function
$F[ {3\over 2}-\nu , {3\over 2}+\nu ; 2; z]$
has a branch point at $z=0$ and a
cut extends from $z=+1$ along the real axis to $z=+\infty .$
The infinitesimal imaginary part of $I$ provides a compact notation for
indicating on which side of the cut the hypergeometric function should
be evaluated. We shall for the most part work in units where
$H^2=1$, and restore $H$ by dimensional analysis later.

To simplify the algebra, we shall in this paper only
consider the special case where
$m^2=2$,  for which the Wightman function has the particularly
simple form
$$
G^{(+)}(X, X^\prime )={-1\over 8\pi ^2}\cdot
{1\over I(X,X^\prime )-1}.
\eqn\jjc$$
In this case the spatial mode functions $S(\sigma ;\zeta )$ also have a
particularly simple form
$$ S(\sigma ;\zeta )=
{1\over \sin [\sigma ]}\exp
\left[ i\zeta \tanh ^{-1}\bigl(\cos \sigma \bigr) \right] =
{e^{i\zeta u}\over \sech [u]} \eqn\jjd$$
where $\tanh [u]=\cos [\sigma ].$
To normalize our mode functions, we use the nondegenerate
bilinear form
$$ (f, g)=(-i)\int _\Sigma d\Sigma ^\mu ~
\biggl\{ f(X)[\partial _\mu g(X)]-
[\partial _\mu f(X)] g(X) \biggr\} \eqn\jje$$
where $\Sigma $ is a Cauchy surface, with unit normal $n^\mu$, and
$d \Sigma^\mu = d\Sigma n^\mu$, with $d\Sigma$ the volume element
on $\Sigma$.
For the $s$-wave, the normalized mode functions are
$$f^{(\pm )}(u, \tau ;\zeta )={1\over 4\pi \sqrt{\zeta }}\cdot
{e^{i\zeta u}\over \sech [u]}\cdot
{e^{\mp i\vert \zeta \vert \tau }\over \cosh [\tau ]},
\eqn\jjf$$
which satisfy the relations
$$\eqalign{
(f^{(+)}_\zeta , f^{(-)}_{\zeta ^\prime })&=+\delta (\zeta -
\zeta ^\prime ),\cr
(f^{(-)}_\zeta , f^{(+)}_{\zeta ^\prime })&=-\delta (\zeta -
\zeta ^\prime ),\cr
(f^{(+)}_\zeta , f^{(+)}_{\zeta ^\prime })&=0,\cr
(f^{(-)}_\zeta , f^{(-)}_{\zeta ^\prime })&=0.\cr }\eqn\jjg$$
We choose this normalization so that when we expand
the s-wave component of the field operator as
$$\hat \phi (X)=\int _{-\infty }^{+\infty }d\zeta \left[
f^{(+)}(X;\zeta )\hat a(\zeta ) +
f^{(-)}(X;\zeta )\hat a^\dagger (\zeta )
\right] ,\eqn\jjh$$
the usual commutation relations
$[\hat a(\zeta ), \hat a^\dagger (\zeta ^\prime )]=
\delta (\zeta -\zeta ^\prime ),$
$[\hat a(\zeta ), \hat a(\zeta ^\prime )]=0,$
and $[\hat a^\dagger (\zeta ), \hat a^\dagger (\zeta ^\prime )]=0$
result.

To expand $G^{(+)}$ in terms of the region II mode
functions, we calculate the matrix elements
$$M(\zeta _1;\pm _1\vert \zeta _2; \pm _2)=
f^{(\pm _1)}(X;\zeta _1)
\circ _{X} G^{(+)}(X,Y) \circ _{Y}
f^{(\pm _2)}(Y;\zeta _2)\eqn\jji$$
where the contraction $\circ $ is defined
according to eqn. \jje . We compute both products
choosing the surface defined by $\tau =0$
to be the Cauchy surface $\Sigma .$

In terms of the region II hyperbolic coordinates,
$$\eqalign{I&(X,X')=
\cos [\sigma ]\cos [\sigma ']\cr
&~+\sin [\sigma ]\sin [\sigma ']
\left( \cosh [\tau ]\cosh [\tau ']\cos [\Theta ]
-\sinh [\tau ]\sinh [\tau '] \right)
-i\epsilon ~\varepsilon (\tau -\tau ') \cr
&=\tanh [u]\tanh [u']\cr
&~~+\sech [u]\sech [u']
\left( Z\cosh [\tau ]\cosh [\tau ']
-\sinh [\tau ]\sinh [\tau '] \right) \cr
&-i\epsilon ~\varepsilon (\tau -\tau ') \cr }\eqn\jjj$$
where
$Z=\cos \Theta [\theta ,\phi ;\theta ',\phi ']=
\cos \theta \cos \theta '+
\sin \theta \sin \theta '\cos (\phi -\phi ').$

Consequently,
$$\eqalign{&G^{(+)}(X,X')
={-1\over 8\pi ^2}\cr
&\times \Bigl[ \tanh [u]\tanh [u']+ \sech [u]\sech [u']
\bigl( Z\cosh [\tau ]\cosh [\tau ']
-\sinh [\tau ]\sinh [\tau '] \bigr) \cr
&~~-1
-i\epsilon ~\varepsilon (\tau -\tau ') \Bigr] ^{-1}\cr
&={-1\over 8\pi ^2}\cdot {1\over
\sech [u]\sech [u']\cosh [\tau ]\cosh [\tau' ]}
\cr
&\times {1\over
Z-\cosh [u-u']\sech [\tau ]\sech [\tau ']-
\tanh [\tau ]\tanh [\tau ']
-i\epsilon ~\varepsilon (\tau -\tau ')}.\cr }\eqn\jjk$$

We evaluate
$$\eqalign{&G^{(+)}(X',X)\circ _Xf^{(\pm )}(X;\zeta )\cr
&=(-i)\int _{-\infty }^{+\infty }du~\sech ^3[u]~ (2\pi )
\int _{-1}^{+1}dZ~{-1\over 4\pi ^2}~
{1\over \sech [u]\sech [u']\cosh [\tau ]\cosh [\tau ']}\cr
&\times {1\over Z-\cosh [u-u']\sech [\tau ]\sech [\tau ']-
\tanh [\tau ]\tanh [\tau '] -i\epsilon ~\varepsilon(\tau ') }\cr
&\times \left( {1\over \sech [u]}\dpartial _\tau \right)
\times \left(
{1\over 4\pi \sqrt{\vert \zeta \vert }}~
{e^{i\zeta u}\over \sech [u]}~
{e^{\mp i\vert \zeta \vert \tau}\over \cosh [\tau ]}
\right) \cr }\eqn\jjl$$
at  $\tau =0.$ (Note primed and unprimed indices have been
interchanged.)
Define
$$\eqalign{
&H(u', \tau ',\tau )=\int _{-1}^{+1}dZ\int _{-\infty }^{+\infty }
du~e^{+i\zeta u}\cr
&\times {1\over
Z-\cosh [u-u']\sech [\tau ]\sech [\tau ']-
\tanh [\tau ]\tanh [\tau ']
-i\epsilon ~\varepsilon (\tau ')
}\cr
&=\int _{-\infty }^{+\infty }du ~
e^{+i\zeta u}
{\rm ln}\left[
{ \cosh [u-u']\sech [\tau ]\sech [\tau ']+
\tanh [\tau ] \tanh [\tau ']-1
+i\epsilon ~\varepsilon (\tau ')\over
\cosh [u-u']\sech [\tau ]\sech [\tau ']+
\tanh [\tau ] \tanh [\tau ']+1
+i\epsilon ~\varepsilon (\tau ') }
\right] \cr
&=\int _{-\infty }^{+\infty }du ~
e^{+i\zeta u}
{\rm ln}\left[
{ \cosh [u-u']-\cosh [\tau '-\tau ]
+i\epsilon ~\varepsilon (\tau ')\over
\cosh [u-u']+\cosh [\tau '+\tau ']
+i\epsilon ~\varepsilon (\tau ') }
\right] \cr } \eqn\jjm$$
so that
$$\eqalign{G^{(+)}(X',X)&\circ _Xf^{(\pm )}(X;\zeta )\cr
&={+i\over 16\pi ^2\sqrt{\vert \zeta \vert }}
{1\over \sech [u']\cosh [\tau ']} \left[
{-\partial H(u',\tau ',\tau )\over \partial \tau }\mp i\vert \zeta \vert
H(u',\tau ',\tau ) \right] \cr }\eqn\jjma$$
at $\tau =0.$

We now evaluate
$$\eqalign{ &\int _{-\infty}^{+\infty }du~
e^{+i\zeta u}~{\rm ln}
\left[
{ \cosh [u-u']-\cosh [\tau  '-\tau ]
+i\epsilon ~\varepsilon (\tau ')\over
\cosh [u-u']+\cosh [\tau '+\tau ]
+i\epsilon ~\varepsilon (\tau ') }
\right] \cr
&= e^{+i\zeta u'}
\int _{-\infty}^{+\infty }du~
e^{+i\zeta u}~
{\rm ln}
\left[
{ \cosh [u]-\cosh [\tau '-\tau ]
+i\epsilon ~\varepsilon (\tau ')\over
\cosh [u']+\cosh [\tau '+\tau ]
+i\epsilon ~\varepsilon (\tau ') }
\right] .\cr }
\eqn\jjo$$

Define
$$I_2^{(\pm )}=
\int _{-\infty}^{+\infty }du~
e^{+i\zeta u}~{\rm ln}
\left[ {
\cosh [u]-\cosh [\tau '-\tau ]
\pm i\epsilon
\over
\cosh [u]+\cosh [\tau '+\tau ]
\pm i\epsilon
}\right] .  \eqn\jjp$$
We first evaluate $I^{(+)}_2.$
Examining the properties of the integrand in the complex
plane, we find two branch points near the real axis, just
above or below the two points $u=\pm (\tau '-\tau ).$
The branch point to the left at
$u=-\vert \tau '-\tau \vert +i\epsilon $
lies just above the real axis and the
branch point at $u=+\vert \tau '-\tau \vert -i\epsilon $
lies just below the real
axis. [Under the transformation $+i\epsilon \to -i\epsilon $
this situation is reversed.]

We now consider the contour $C$
from $-\bar T$ to $+\bar T$ to $\bar T+2\pi i$ to
$-\bar T+2\pi i$ and finally back to $-\bar T,$
considered in the limit $\bar T\to \infty .$
Inside the rectangle enclosed by the contour,
in addition to the branch point
at $u=-\vert \tau '-\tau \vert +i\epsilon ,$
there are three more branch points
at $u=-\vert \tau '+\tau \vert +i\pi -i\epsilon ,$
$u=+\vert \tau '+\tau \vert +i\pi +i\epsilon ,$
and $u=+\vert \tau '-\tau \vert +i2\pi -i\epsilon .$
One branch cut connects
$u=-\vert \tau '-\tau \vert +i\epsilon $ to
$u=-\vert \tau '+\tau \vert +i\pi -i\epsilon ,$ and
another branch cut connects
$u=+\vert \tau '+\tau \vert +i\pi +i\epsilon $ to
$u=+\vert \tau '-\tau \vert +i2\pi -i\epsilon .$
In the limit $\bar T\to \infty $
the contributions from the vertical parts of the contours
vanish. Since a translation by $2\pi i$ has the effect of
multiplying the integrand by $e^{- 2 \zeta \pi },$
$$\eqalign{
\oint _C& du~
e^{+i\zeta u}~{\rm ln}
\left[ {
\cosh [u]-\cosh [\tau '-\tau ]
+i\epsilon
\over
\cosh [u]+\cosh [\tau '+\tau ]
+ i\epsilon
 }\right] \cr
&=(1-e^{-2\zeta \pi })~I^{(+)}\cr
&=(-2\pi i)\int _{-\vert \tau '-\tau \vert }^%
{-\vert \tau '+\tau \vert +i\pi }du~e^{+i\zeta u}
+( 2\pi i)\int _{+\vert \tau '+\tau \vert +i\pi }^%
{+\vert \tau '-\tau \vert +2i\pi }du~e^{+i\zeta u}\cr
&={2\pi \over \zeta }
\left[
(e^{-i\zeta \vert \tau '-\tau \vert }
-e^{-\zeta \pi -i\zeta  \vert \tau '+\tau \vert })
-(e^{-\zeta \pi +i\zeta \vert \tau '+\tau \vert }-
e^{-2\zeta \pi +i\zeta \vert \tau '-\tau \vert })
\right]
\cr }\eqn\jjq$$
where the integral is evaluated by deforming the
contour into two pieces surrounding the two branch
cuts. It follows that
$$\eqalign{ I^{(+)}&=-{2\pi \over \zeta \sinh [\zeta \pi ]}
\biggl[ \cos [\zeta  \vert \tau '+\tau \vert ]-\cosh [\zeta \pi -i\zeta
 \vert \tau '-\tau \vert ] \biggr] \cr
&=-{2\pi \over \zeta \sinh [\zeta \pi ]}
\biggl[ (
\cos [\zeta  \vert \tau '+\tau \vert ]
-\cosh [\zeta \pi ])\cos [\zeta  \vert \tau '-\tau \vert ]\cr
&\hskip 25pt
+i\sinh [\zeta \pi ]\sin [\zeta  \vert \tau '-\tau \vert ]
\biggr] .\cr }\eqn\jjr$$
Similarly,
$$\eqalign{ I^{(-)}&=-{2\pi \over \zeta \sinh [\zeta \pi ]}
\biggl[ \cos [\zeta  \vert \tau '+\tau \vert ]-\cosh [\zeta \pi +i\zeta
 \vert \tau '-\tau \vert ] \biggr] \cr
&=-{2\pi \over \zeta \sinh [\zeta \pi ]}
\biggl[ (
\cos [\zeta  \vert \tau '+\tau \vert ]
-\cosh [\zeta \pi ])\cos [\zeta  \vert \tau '-\tau \vert ]\cr
&\hskip 25pt
-i\sinh [\zeta \pi ]\sin [\zeta  \vert \tau '-\tau \vert ]
\biggr] .\cr }
\eqn\jjs$$

It follows for both signs of $\tau '$ that
$$\eqalign{H&=-{2\pi \over \zeta \sinh [\zeta \pi ]} e^{+i\zeta u'}
\biggl\{ (\cos [\zeta (\tau '+\tau )]
-\cosh [\zeta \pi ])\cos [\zeta (\tau '-\tau )]\cr
&~~~~
+i\sinh [\zeta \pi ]\sin [\zeta (\tau '-\tau )] \biggr\} .\cr }
\eqn\jjt$$
Therefore
$$\eqalign{
(G^{(+)}&\circ f^{(\pm )}_\zeta )(X')=
{-i\over 8\pi \sqrt{\vert \zeta \vert }}
{1\over \sinh [\zeta \pi ]}
{e^{+i\zeta u'}\over \sech [u']\cosh [\tau ']}\cr
&\times
\left( {-1\over \zeta }{\partial \over \partial \tau }
\mp i{\vert \zeta \vert \over \zeta } \right) \cr
&\times
\left[ \cos [\zeta (\tau '+\tau )]
-\cosh [\zeta \pi ]\cos [\zeta (\tau '-\tau )]
+i\sinh [\zeta \pi ]\sin [\zeta (\tau '-\tau )]
\right] .\cr }
\eqn\jju$$
Since
$$\eqalign{
&\left( {1\over \zeta }{\partial \over \partial \tau '}
\mp i{\vert \zeta \vert \over \zeta } \right)
\cos [\zeta \tau ']=
\mp i {\vert \zeta \vert \over \zeta }~
e^{\mp i\vert \zeta \vert \tau '},\cr
&\left( {1\over \zeta }{\partial \over \partial \tau '}
\mp i{\vert \zeta \vert \over \zeta } \right) \sin [\zeta \tau ']
=e^{\mp i\vert \zeta \vert \tau '},\cr
}\eqn\jjv$$
eqn. \jju ~ may be rewritten (setting $\tau=0$) as
$$\eqalign{
&(G^{(+)}\circ f^{(\pm )}_\zeta )(X')=
{1\over 8\pi \sqrt{\vert \zeta \vert }}
{e^{+i\zeta u'}
\over \sech [u']\cosh [\tau ']}\cr
&~\times {-1\over \sinh [\zeta \pi ]}
\left[
-e^{\mp i\vert \zeta \vert \tau '}
\sinh [\zeta \pi ]\mp {\vert \zeta \vert \over \zeta }\cdot
\bigl\{ -e^{\pm i\vert \zeta \vert \tau '}
+e^{\mp i\vert \zeta \vert \tau '}\cosh [\zeta \pi ]\bigr\}
\right] \cr
&={1\over 4\pi \sqrt{\vert \zeta \vert }}
{e^{+i\zeta u'}~e^{\mp i\vert \zeta \vert \tau '}
\over \sech [u']\cosh [\tau ']}\cr
&~~~~\times
{-1\over 2\sinh [\vert \zeta \vert \pi ]}
\cdot \biggl\{
\mp e^{+\vert \zeta \vert \pi }e^{\mp i\vert \zeta \vert \tau '}
\pm e^{\pm  i\vert \zeta \vert \tau '} \biggr\}.
\cr }
\eqn\jjw$$
Therefore
$$\eqalign{
G^{(+)}\circ
&
\pmatrix{ f^{(+)}_\zeta \cr f^{(-)}_\zeta \cr }=(-P^{(+)})\cdot
\pmatrix{ f^{(+)}_\zeta \cr f^{(-)}_\zeta \cr }\cr
&={-1\over e^{+\vert \zeta \vert \pi } -
e^{- \vert \zeta \vert \pi }}\cdot
\pmatrix{e^{+\vert \zeta \vert \pi }&-1\cr
+1&-e^{-\vert \zeta \vert \pi }\cr }
\pmatrix{ f^{(+)}_\zeta \cr f^{(-)}_\zeta \cr } ,\cr }
\eqn\jjx$$
and the normalized positive and negative frequency
mode functions are
$$\eqalign{
g^{(+)}&= { e^{\vert \zeta \vert \pi /2} f^{(+)}-
e^{-\vert \zeta \vert \pi /2} f^{(-)}
\over  \left( e^{+\vert \zeta \vert \pi } -
e^{- \vert \zeta \vert \pi }
\right) ^{1/2} },\cr
g^{(-)}&= { e^{\vert \zeta \vert \pi /2} f^{(-)}-
e^{-\vert \zeta \vert \pi /2} f^{(+)}
\over  \left( e^{+\vert \zeta \vert \pi } -
e^{- \vert \zeta \vert \pi }
\right) ^{1/2} }.\cr
}\eqn\jjy$$
{}From the fact that $G^{(+)}$ obeys the wave equation
everywhere in both variables and from the orthonormality
relations for the mode functions given in eqn. \jjh ,
it follows that
$$\eqalign{
G^{(+)}(X,X')&=\int _{-\infty }^{+\infty }d\zeta ~
\sum _{\pm _1}\sum _{\pm _2}\cr
&\times (\pm _1)~(\mp _2) ~\biggl( f^{(\pm _1)}_\zeta (Y)~
\circ _Y G^{(+)}(Y,Y') \circ _{Y'}~ f^{(\pm _2)}_\zeta (Y')
\biggr) \cr
& \times
f^{(\pm _1)}_\zeta (X)~ f^{(\pm _2)}_\zeta (X')\cr }
\eqn\jjz$$
The matrix elements
$f^{(\pm _1)}_\zeta (Y)~
\circ _Y G^{(+)}(Y,Y') \circ _{Y'}~ f^{(\pm _2)}_\zeta (Y')$
are given in eqn.  \jjx .

\def\H{{\cal H}}
\def\K{{\cal K}}

\chapter{Cosmological Perturbations}

In the previous sections, we calculated the
quantum fluctuations of the inflaton field $\phi (x)$
about the classical background solution $\phi _{b}(x)$
treating the curved space as fixed background, determined
by the background solution with no fluctuations. To ignore
the back reaction of the linear perturbations on the
background spacetime is a valid approximation when the
perturbations are well within the Hubble radius. However, as the
expansion of the universe causes the size of a
perturbation to increase relative to the Hubble length, this
approximation breaks down. In this section we consider the
coupling to gravity of the scalar field perturbations---in
other words, how the quantum fluctuations in the inflaton
field translate into density perturbations as they are pushed
outside the Hubble radius. It is permissible to treat the
fluctuations in the scalar field (and their associated metric
perturbations) as classical fields.

There are many physically equivalent ways to treat density
perturbations. Here we use the gauge-invariant formalism of
Bardeen, \refmark{\bardeen } closely following the notation in the
review in ref. \mfb . We consider only scalar perturbations and
write the metric in the form
$$ds^2=a^2(\eta )\cdot
\biggl[
-(1+2\varphi )d\eta ^2+2B_{\vert i}~d\eta ~dx^i
+\bigl\{  (1-2\psi )\gamma _{ij} +2E_{\vert ij}\bigr\}
dx^i~dx^j\biggr] \eqn\yyya$$
where $\eta $ is conformal time, and the spacetime functions
$\varphi ,$ $\psi ,$ $B$ and $E$ are linear metric perturbations.
Primes denote derivatives with
respect to conformal time and $\H =a'/a=aH.$ Covariant
spatial differentiation with respect to the spatial background
metric $\gamma_{ij}$ is denoted $D_j F = F_{\vert j}$ etc.

We set
$$\gamma _{ij}~dx^i~dx^j=
d\xi ^2+f^2(\xi )d\Omega ^2_{(2)}\eqn\yyac$$
where
$$ f(\xi )=\cases{ \sinh [\xi ],
&for $\K =-1,$\cr \xi , &for $\K =0,$\cr
\sin [\xi ], &for $\K =+1.$\cr } \eqn\yyab$$
Although we are primarily interested in the open geometry
$(\K =-1),$ we consider the cases $\K =-1,$ $0,$ and $+1,$
(corresponding to open, flat, and closed spatial geometries,
respectively) to allow comparison between the open and flat
cases, so that the influence of spatial
curvature is manifest.\footnote\dagger{ In ref.
\mfb ~ the metric is written as
$(dr^2+r^2d\Omega ^2_{(2)})/(1+\K r^2/4)^2.$ The substitutions
$r=2\tanh [\xi /2]$ for $\K =-1$ and $r=2\tan [\xi /2]$ for $\K =+1$
demonstrate equivalence to the line element in eqn. \yyac .
Likewise, in
ref. \kodama ~ the line element is written as
$d\bar r^2/(1-\K \bar r^2)+\bar r^2d\Omega ^2_{(2)}.$ Equivalence
to eqn.  \yyac ~ may be demonstrated by the substitutions
$\bar r=\sinh [\xi ]$ and $\bar r=\sin [\xi ]$ for $\K =-1$
and  $\K =+1,$ respectively.}
In comparing open and flat geometries, it should be kept in
mind that spatial curvature affects the eigenvalues of the Laplacian
$\nabla ^2=\gamma ^{ij}D_iD_j$ used in the mode decomposition. For $\K =0,$
the eigenvalues of $-\nabla ^2$ are $k^2$ where $k$ ranges from $0$
to $+\infty ;$ for $\K =-1,$ the eigenvalues of $-\nabla ^2$ are
$\zeta ^2+1$ where $\zeta $ ranges from $0$ to $+\infty .$

{}From these linear perturbations, we construct the
`gauge-invariant' variables
$$\eqalign{
\Phi &=\varphi +{1\over a}~\bigl[ (B-E')~a\bigr] ^\prime  ,\cr
\Psi &=\psi -{a'\over a}~[B-E'],\cr
}\eqn\yyb $$
where `gauge-invariant' means invariant to linear order
under infinitesimal coordinate transformations.
We construct the `gauge-invariant' perturbations of the
Einstein tensor
$$\eqalign{
\delta {\cal G}_0^0&= \delta G_0^0+[~{}^bG_0^0]'~(B-E')\cr
\delta {\cal G}_i^0&= \delta G_i^0+[~{}^bG_i^0-{1\over 3} ~{}^bG_k^k ]~
(B-E')_{\vert i}\cr
\delta {\cal G}_j^i&= \delta G_j^i+[~{}^bG_j^i]'~(B-E'),\cr
}\eqn\yyc$$
and similarly of the stress-energy tensor
$$\eqalign{
\delta {\cal T}_0^0&= \delta T_0^0+[~{}^bT_0^0]'~(B-E'),\cr
\delta {\cal T}_i^0&= \delta T_i^0+[~{}^bT_i^0-{1\over 3} ~{}^bT_k^k ]~
(B-E')_{\vert i},\cr
\delta {\cal T}_j^i&= \delta T_j^i+[~{}^bT_j^i]'~(B-E').\cr
}\eqn\yyca$$

In terms of the gauge-invariant metric perturbations,
$$\eqalign{\delta {\cal G}_0^0&=
{2\over a^2}\bigl[ -3\H (\H \Phi +\Psi ')+\nabla ^2\Psi +3\K \Psi \bigr] ,\cr
\delta {\cal G}_i^0&=
{2\over a^2}\bigl[
\H \Phi +\Psi '\bigr] _{,i}~,\cr
\delta {\cal G}_j^i&=
{-2\over a^2}\bigl[ ~\bigl\{
(2\H '+\H ^2)\Phi +\H \Phi '+\Psi ^{\prime \prime }+2\H\Psi '-\K \Psi
+{1\over 2}\nabla ^2D\bigr\}\delta ^i_j\cr
&\hskip 1in   -{1\over 2}\gamma ^{ik}D_{\vert kj } \bigr] \cr
}\eqn\yyd$$
where $D=(\Phi -\Psi ).$
The gauge-invariant stress-energy from the inflaton field is
$$ \eqalign{
\delta {\cal T}_0^0&=
{1\over a^2}\left[ -\phi _b^{\prime ~2}\Phi +\phi _b^\prime ~
(\delta \hat \phi )^{\prime }
+a^2V_{,\phi }~(\delta \hat \phi )\right]\cr
\delta {\cal T}_i^0&= {1\over a^2}\left[  \phi _b^\prime ~
(\delta \hat \phi  )\right]_{,i}\cr
\delta {\cal T}_j^i&=
{1\over a^2}\left[ +\phi _b^{\prime ~2}\Phi -\phi _b^\prime ~
(\delta \hat \phi )^{\prime }
+a^2V_{,\phi }~(\delta \hat \phi )\right] \delta ^i_j\cr
}\eqn\yye$$
where
$(\delta \hat \phi )=(\delta \phi )
+\phi '_b~(B-E')$ is the gauge-invariant variable
for the inflaton field perturbation.
The linearized Einstein equation
$\delta {\cal G}_\mu ^\nu =(8\pi G)~\delta {\cal T}_\mu ^\nu $
gives $D=0$ (so that $\Psi =\Phi $) and the equations
$$\eqalign{
3(\K -\H ^2)\Phi -3\H \Phi '+\nabla ^2\Phi  &=(4\pi G)
\bigl[ -\phi _b^{\prime ~2}\Phi  +
\phi _b^\prime ~(\delta \hat \phi )^{\prime }
+a^2V_{,\phi }~(\delta \hat \phi )\bigr] ,\cr
\H \Phi +\Phi ' &=(4\pi G)
\bigl[   \phi _b^{\prime }~(\delta \hat \phi )\bigr] ,\cr
\Phi ^{\prime \prime }+3\H \Phi '+(2\H ^\prime +\H ^2-\K )\Phi
&=(4\pi G)\cdot \bigl[  -\phi _b^{\prime ~2}\Phi
+\phi _b^\prime ~(\delta \hat \phi )^{\prime }
-a^2V_{,\phi }~(\delta \hat \phi ) \bigr] .\cr }\eqn\yyf$$

Because of the background Einstein equations
$$\eqalign{
&{3\over a^2}\bigl[ \H ^2+\K \bigr] =(8\pi G)\cdot (+\rho _b)=
(8\pi G)\cdot
\left[ {1\over 2a^2}\phi _b^{\prime ~2}+V(\phi _b)\right] ,\cr
&{1\over a^2}\bigl[ 2\H ^\prime +\H ^2 +\K \bigr] =
(8\pi G)\cdot (-p_b)= (8\pi G)\cdot
\left[ -{1\over 2a^2}\phi _b^{\prime ~2}+V(\phi _b)\right] ,\cr }
\eqn\yyh$$
it follows that $(4\pi G)\phi _b^{\prime ~2}=
(\H ^2-\H ^\prime +\K ).$ Thus eqn. \yyf ~ may be rewritten as
$$\eqalign{-3\H \Phi '+\nabla ^2\Phi +(4\K -\H ^\prime -2\H ^2)\Phi
&=(4\pi G)\cdot
\bigl[ \phi _b^\prime ~(\delta \hat \phi )^{\prime }
+a^2V_{,\phi }~(\delta \hat \phi )\bigr] ,\cr
\H \Phi +\Phi ' &=(4\pi G)\cdot
\bigl[   \phi _b^{\prime }~(\delta \hat \phi )\bigr] ,\cr
\Phi ^{\prime \prime }+3\H \Phi '+(\H '+2\H ^2)\Phi
&=(4\pi G)\cdot \bigl[
\phi _b^\prime ~(\delta \hat \phi )^{\prime }
-a^2V_{,\phi }~(\delta \hat \phi ) \bigr] .\cr }\eqn\yyi$$

We now subtract the first equation from the third equation. Because
of the background equation of motion for the scalar field
$$ \phi _b ^{\prime \prime }
+2\H \phi _b^\prime +a^2V_{,\phi }( \phi _b)=0, \eqn\yyj$$
we can subtract $\bigl[ 4\H -2(\phi _b^{\prime \prime }/\phi _b^\prime )
\bigr] $
times the second equation, obtaining
$$ \Phi ^{\prime \prime }+2\left( \H -
{\phi _b^{\prime \prime }\over \phi _b^\prime }\right)
\Phi ^\prime -\nabla ^2\Phi +\left(
2\H ^\prime -2\H {\phi _b^{\prime \prime }\over \phi _b^\prime }-4\K
\right) \Phi =0 .\eqn\yyk $$
We rewrite eqn. \yyk ~ in terms of proper time, using the
relation $dt=a(\eta )d\eta ,$ so that
$$\ddot \Phi +\left( H-2{\ddot \phi _b\over \dot \phi _b}\right)
\dot \Phi +
{1\over a^2}\left( -\nabla ^2-4\K
\right) \Phi
+2\left( \dot H-H
{\ddot \phi _b\over \dot \phi _b}
\right)  \Phi =0.\eqn\yyl$$

We now relate the gauge invariant potential perturbations to the
gauge invariant scalar field perturbations. From eqns. \yyi ~  and
\yyk ~ it follows that
$$\eqalign{ (\delta \hat \phi )&=
{1\over (4\pi G)}\cdot {1\over \phi _b^\prime }
\left[ \Phi ^\prime +\H \Phi \right] ,\cr
(\delta \hat \phi )^\prime &=
{-1\over (4\pi G)}\cdot {1\over \phi _b^\prime } \left[
\left( \H -{\phi _b^{\prime \prime }
\over \phi ^\prime _b}\right)\Phi ^\prime
+\left( \zeta ^2+5+\H ^\prime -\H
{\phi _b^{\prime \prime }\over \phi ^\prime _b}\right)\Phi
\right] ,\cr }\eqn\bdgb$$
so that in terms of proper time derivatives
$$\eqalign{ (\delta \hat \phi )&=
{1\over (4\pi G)}\cdot {1\over \dot \phi _b}
\left[ \dot \Phi  +H \Phi \right] ,\cr
\partial _t(\delta \hat \phi )&=
{-1\over (4\pi G)}\cdot {1\over \dot \phi _b } \left[
\left( -{\ddot \phi _b\over \dot \phi _b}\right)\dot \Phi
+\left\{ {\zeta ^2+5\over a^2}+
\dot \phi _b ~\partial _t\left( {H\over \dot \phi _b}\right)
\right\}  \Phi
\right] .\cr }\eqn\bdga$$

To calculate the evolution of the density perturbations during
the early part of region I inflation (until $\Omega $ is close to one),
we use a linear approximation to the potential. In the general case,
at $t=0$ (at the end of the Coleman--de Luccia bounce),
$\dot \phi _b=0$ and the potential slopes downward, toward
the true vacuum. To make the problem tractable analytically, we
take
$$V(\phi )=V_0-V _{,\phi }\phi ,$$
where we set $\phi _b(t=0)=0$ and take $V _{,\phi }$ to be
constant.
We first solve for the evolution of $\dot \phi _b,$ for simplicity
assuming that $a(t)$ is well approximated by $a(t)=\sinh [t]$ until
$\Omega $ is close to one. Thus the equation of motion for $\phi _b$
is
$$\ddot \phi _b+3\coth [t]~\dot \phi _b=V_{,\phi },\eqn\yys$$
and the solution with $\dot \phi _b=0$ at $t=0$ is
$$
\dot \phi _b(t)=V_{,\phi }\cdot
{\cosh ^3[t]-3\cosh [t]+2
\over 3\sinh ^3[t]},\eqn\yyr$$
which for small $t$ behaves as ${1\over 4}V_{,\phi }t$
and for large $t$ as ${1\over 3}V_{,\phi }.$
It follows that
$${\ddot \phi _b(t)\over \dot \phi _b(t)}=
{3\over \sinh [t]\cdot (\cosh [t]+2 )},\eqn\yypaa$$
so that eqn. \yyl ~ becomes
$$\eqalign{&\ddot \Phi +
\left\{ {1\over \sinh [t]}\left( \cosh [t] - {6\over \cosh [t]+2}
\right) ~
\right\} \dot \Phi \cr
&\hskip 70pt +{1\over \sinh ^2[t]}\left\{ \zeta ^2+5-
{4\cdot (2\cosh [t]+1)\over \cosh [t]+2}\right\} \Phi =0.\cr }
\eqn\yyp$$
To solve the eqn. \yyp ~
it is convenient to replace the variable $t$ with the
conformal time variable
$\eta ={\rm ln}\bigl[ \tanh [t/2]\bigr] ,$ thus mapping the interval
$(0<t<+\infty )$ into $(-\infty <\eta <0),$ so that eqn. \yyp ~
becomes
$$\Phi ^{\prime \prime }-{6(1-e^{2\eta })\over 3-e^{2\eta }}\Phi ^\prime
+\left[ (\zeta ^2+5)-{4(3+e^{2\eta })\over 3-e^{2\eta }}\right] \Phi
=0.\eqn\appbc$$
The general solution to eqn. \appbc ~ is\footnote\dagger{We thank
Bharat Ratra for pointing out that eqn. \appbc ~ can be solved
analytically.}
$$\eqalign{
\Phi =
&c_{(+)}~\left[e^{+i\zeta \eta } \cdot e^\eta \cdot
\left( 1-{(\zeta +i)\over 3(\zeta -i)}e^{2\eta } \right) \right] \cr
+ & c_{(-)}~\left[ e^{-i\zeta \eta } \cdot e^\eta \cdot
\left( 1-{(\zeta -i)\over 3(\zeta +i)}e^{2\eta }
\right) \right] .\cr }
\eqn\rvva $$
For small $t$ (i.e., $\eta \to -\infty $), $t\approx 2e^\eta $
and $\Phi \approx c_{(+)}e^\eta e^{+i\zeta \eta }
+c_{(-)}e^\eta e^{-i\zeta \eta }=
c_{(+)}(t/2)(t/2)^{+i\zeta }+c_{(-)}(t/2)(t/2)^{-i\zeta }.$
For large $t$ (i.e., $\eta \to 0-$), $\Phi $ is dominated by
the growing mode and
$$
\Phi \approx
c_{(+)} {2\over 3}\left( {\zeta +2i\over \zeta -i}\right) +
c_{(-)} {2\over 3}\left( {\zeta -2i\over \zeta +i}\right) .
\eqn\rvvb $$

Finally, inserting the small $t$ asymptotic forms
$\Phi \sim t^{\pm i\zeta +1},$
$a\approx t,$
$H\approx t^{-1},$
$\dot \phi _b\approx {1\over 4}V_{,\phi }t,$
and
$\ddot \phi _b\approx {1\over 4}V_{,\phi }$
into eqn. \bdga , we obtain the small $t$ matching condition
$$
(\delta \hat \phi )\approx {(\pm i\zeta +2)\over \pi GV_{,\phi }}
\times {\Phi \over t^2}.\eqn\bdgc$$

In spatially flat inflation the conserved quantity
$$\chi ={2\over 3}{\H ^{-1}\Phi ^\prime +
\Phi \over 1+w}+\Phi \eqn\zinv$$
[called $\zeta $ in ref. \mfb ~]
is very useful to track density perturbations on
super-Hubble radius scales. During inflation
$$w=p_b/\rho _b=(-1)\cdot {V[\phi _b]-{1\over 2}\dot \phi ^2_b\over
V[\phi _b]+{1\over 2}\dot \phi ^2_b } \approx
{1\over 24\pi }\left( {m_{pl}V_{,\phi }\over V}\right) ^2-1.
\eqn\yyya$$
Assuming the absence of entropy
perturbations (i.e., that the perturbations obey the
equation of state $p_b(\rho _b)$ given by the background
solution) and neglecting the spatial derivative term in the evolution equation
for the gauge-invariant potential $\Phi ,$ one obtains $\dot \chi =0.$
On super-Hubble radius scales these assumptions hold reasonably well. In flat
inflation the existence of this conserved quantity provides an elegant
way of demonstrating that the spectrum of density perturbations
at late times is
independent of the details of re-heating. The precise
way in which $w$ evolves, from
a value slightly greater than $-1$ during inflation to ${1\over 3}$ during
radiation domination, and later to nearly zero during matter domination,
does not affect the final density perturbations.

Unfortunately, the conservation of $\chi $ for super-Hubble radius
 scales does not
completely generalize to an open expanding universe. $\chi $ varies with
time when $\Omega $ is not close to one, in our scenario
during the early part of region I inflation and then much later, during
the latter part of the matter domination, when the universe becomes
curvature dominated again.  Therefore we may use $\chi $ in the following way.
During the early part of inflation (when $\Omega $ is significantly less
than one), we calculate the evolution of the modes explicitly. Then at a
later time (before reheating but
when $\Omega $ is still very nearly one), when all of the modes
of interest are well outside the Hubble radius, we calculate $\chi .$
(It should be noted that while $\Omega $ is less than one,
$\chi $ varies in time in a manner independent of
wave number. Therefore, all modes are affected in the same way, and only the
overall normalization of the power spectrum is altered.
Its shape remains the same. This is because the evolution of
density perturbations on super-Hubble radius scales is essentially a local
process.)

\chapter{Power Spectrum of the Inflationary Open Universe}

In the last three sections we developed the tools
for calculating density perturbations. In this section
we put together the various pieces to give a concrete result.
To keep the calculation as simple as possible, we
consider the following idealized scenario. In region II
we assign to the inflaton field perturbations a mass $m^2=2,$
which changes discontinuously across the light cone to
$m^2=0$ in region I. This should be a reasonable approximation
when the bubble wall is very thin and the bubble size is
small compared to the
Hubble length. [The scenarios with two scalar fields
work best for this regime. See the discussion at the end of Section IV.]

We start with the mode functions associated
with the annihilation operators of the Bunch-Davies
initial state in  region II (where $m^2=2$), determined
in eqn. \jjy ~ to be
$$\eqalign{
g^{(+)}_\zeta &= { e^{\vert \zeta \vert \pi /2} f^{(+)}_\zeta -
e^{-\vert \zeta \vert \pi /2} f^{(-)}_\zeta
\over  \left( e^{+\vert \zeta \vert \pi } -
e^{- \vert \zeta \vert \pi }
\right) ^{1/2} },\cr
g^{(-)}_\zeta &= [g^{(+)}_\zeta ]^*,\cr }\eqn\bdgz$$
where
$$
f^{(\pm )}_\zeta (u, \tau )={1\over 4\pi \sqrt{\vert \zeta \vert }}\cdot
{e^{i\zeta u}\over \sech [u]}\cdot
{e^{\mp i\vert \zeta \vert \tau }\over \cosh [\tau ]}.
\eqn\bdgy$$
We continue these mode functions into region I.
For large $u,$ $(e^{i\zeta u}/\sech [u])\approx
(1/\sigma )\cdot (\sigma /2)^{-i\zeta }.$ Therefore
in region II near the light cone one has
$$\eqalign{
f^{(\pm )}_\zeta &\approx
{1\over 4\pi \sqrt{\vert \zeta \vert }}\cdot
{e^{\mp i\vert \zeta \vert \tau }\over \cosh [\tau ]}\cdot
{1\over \sigma }~\left( {\sigma \over 2}\right) ^{-i\zeta }.
\cr } \eqn\sverd$$
Recall that (as shown in appendix A) the matching
conditions across the light cone are
$$\eqalign{
\sigma ^{+i\zeta -1}\cdot
{e^{+i\zeta \tau }\over \cosh [\tau ]}
&\to
(2i)\cdot {\sin [\zeta \xi ]\over \sinh[\xi ]}
\cdot t^{+i\zeta -1},\cr
\sigma ^{+i\zeta -1}\cdot
{e^{-i\zeta \tau }\over \cosh [\tau ]}
&\to 0,\cr
\sigma ^{-i\zeta -1}\cdot
{e^{+i\zeta \tau }\over \cosh [\tau ]}
&\to 0,\cr
\sigma ^{-i\zeta -1}\cdot
{e^{-i\zeta \tau }\over \cosh [\tau ]}
&\to
(-2i)\cdot {\sin [\zeta \xi ]\over \sinh[\xi ]}
\cdot t^{-i\zeta -1}.\cr
}\eqn\sverdf$$
Therefore, for $\zeta >0$ only $f^{(+)}_\zeta $ has a
nonvanishing continuation into region I, and similarly for
$\zeta <0$ only $f^{(-)}_\zeta $  has a nonvanishing
continuation into region I. We set $\zeta >0.$ In region I
near the light cone
$$\eqalign{
g^{(+)}_\zeta &=
{e^{\pi \zeta /2}\over
\sqrt{e^{\pi \zeta }-e^{-\pi \zeta }}}\cdot
{-i\over 2\pi \sqrt{\zeta }} \cdot 2^{+i\zeta }\cdot
{\sin [\zeta \xi ]\over \sinh [\xi ]}\cdot t^{-i\zeta -1}\cdot
[1+O(t)],\cr
g^{(+)}_{-\zeta }&=
{e^{-\pi \zeta /2}\over
\sqrt{e^{\pi \zeta }-e^{-\pi \zeta }}}\cdot
{-i\over 2\pi \sqrt{\zeta }}\cdot
2^{-i\zeta }\cdot {\sin [\zeta \xi ]\over \sinh [\xi ]}\cdot
t^{+i\zeta -1}\cdot
[1+O(t)].\cr
}\eqn\sverdg$$

We now change variables to the gauge-invariant potential
$\Phi $ using eqn. \bdgc , thus introducing a factor of
$t^2~\pi GV_{,\phi }/(\mp i\zeta +2).$ We write $\hat \Phi $ as a
quantum-mechanical operator, so that the s-wave component
$$
\hat \Phi =\int _0^\infty d\zeta ~\left[
\hat \Phi _\zeta ^{(+)}
+\hat \Phi _\zeta ^{(-)}
\right] \cdot {\sin [\zeta \xi ]\over \sinh [\xi ]}
\eqn\sverdga$$
where
$$\eqalign{\hat \Phi _\zeta ^{(+)}&=\pi G~V_{,\phi } \cdot
{-i\over 2\pi \sqrt{\zeta }}\cdot
{1\over \sqrt{e^{\pi \zeta }-e^{-\pi \zeta }}}\cr
&\times \left[
e^{\pi \zeta /2}~{2^{i\zeta}
\over -i\zeta +2}~F(t;\zeta ) ~\hat a_\zeta
+e^{-\pi \zeta /2}~{2^{-i\zeta}\over i\zeta +2}~F(t;\zeta )^* ~\hat a_{-\zeta }
\right] \cr }
\eqn\sverdgb$$
and $F(t;\zeta )$ satisfies eqn. \yyp ~ and is normalized to behave
as $F(t;\zeta )\approx t^{-i\zeta +1}$ for small $t,$
so that
$$
F(t;\zeta )=2^{(1-i\zeta )}~e^{+i\zeta \eta }~e^\eta ~
\left[ 1-{\zeta +i\over 3(\zeta -i)}
\right] .
\eqn\revzz$$
$\hat \Phi ^{(+)}$ is the `positive frequency' part of
$\hat \Phi $---that part
which annihilates the Bunch-Davies vacuum. The expression
for $\hat \Phi _\zeta $ (which consists of one quantum mechanical
degree of freedom) involves two harmonic oscillator degrees of
freedom, rather than just one such degree of freedom, because there
exist correlations between regions I and II. (If one considers
the Bunch-Davies vacuum restricted to region I, one has a
mixed state rather than a pure state because of correlations across
the light cone.)

As discussed in the previous section, $F(t;\zeta )$ oscillates for small
$t,$ while the mode labelled by $\zeta $ is within the Hubble radius, and
then for large $t,$ when the mode is well outside the Hubble radius,
$F(t;\zeta ) \approx C^{(g)}_\zeta +C^{(d)}_\zeta e^{-t}.$
We are primarily interested in the growing mode at late times.
Therefore we compute the power spectrum for $\Phi $ as
$$\eqalign{
\lim _{t\to \infty } &\left< \Phi _\zeta (t)\Phi _{\zeta '}(t) \right> \equiv
P_\Phi (\zeta )\delta (\zeta -\zeta ') \cr
P_\Phi (\zeta )
&=\left( {GV_{,\phi } \over H}\right) ^2\cdot
{4\over 3}\cdot {\coth [\pi \zeta ] \over \zeta (\zeta ^2+1)}.\cr }
\eqn\wca$$
where by dimensions we have restored the Hubble constant $H$ during
inflation. In this formula we maintain $\zeta$ as a dimensionless
variable. In terms of the Laplacian on hyperbolic space, we have
$-\nabla^2 = k^2 = H^2(\zeta^2 +1).$ Using $\chi =16\pi G(V^2/V_{,\phi }^2)
\Phi $ (during inflation), one obtains that the
power spectrum for $\chi $ (which has a normalization
that relates more directly to the density perturbations
seen after reheating) is
$$P_\chi (\zeta )={9\over 4 \pi ^2}\cdot \left( {H^3\over
V_{,\phi }}\right) ^2 \cdot
{\coth [\pi \zeta ]\over \zeta (\zeta ^2+1)}.
\eqn\pspchi$$
With the conventions used here $P\sim \zeta ^{-3}$ corresponds
to scale invariance. This is seen for example by computing
$\langle \Phi^2(0)\rangle $ using the small $\xi$ limit of
eqn. \sverdga ~ and noting that there is a logarithmic divergence
at large $\zeta$.
This result generalizes the standard calculations
\refmark{\iperth, \iperts, \ipertgp, \ipert} of fluctuations
produced in a flat universe during  inflation to our
open inflationary scenario.
We compute the CMB perturbations and other cosmological
consequences in a separate letter.\refmark{\nextpaper }

Our result differs from that of Lyth and
Stewart \refmark{\lyth }, and of Ratra and Peebles
\refmark{\rp, \rptwo}
who assume different initial conditions for the quantum fields,
only by the factor
$\coth [\pi \zeta /2].$ This is very close to unity
on all scales accessible to observation (e.g. in the
large angle CMB anisotropy), so we expect very similar
phenomenology to the Lyth-Stewart-Ratra-Peebles spectrum.
However,
it is probably fair to say that our calculation, in
which the initial conditions are physically justified,  puts
the result on a firmer footing.

\chapter{Concluding Remarks}

Finally, we conclude with the following comments:

1. Although it is possible, as we have shown, to extend
inflation from a theory that predicts $\Omega =1$ to one
that predicts $\Omega \le 1,$ we do not believe that
a similar extension is possible for $\Omega >1.$
To be sure, one can construct a closed inflationary model
by postulating positive spatial curvature at the beginning
of inflation, but then inflation no longer solves the
smoothness problem, and much of the original motivation
for inflation is lost.

The fundamental distinction between closed and open
inflation is that negative curvature can be introduced
{\it locally,} by an event localized in space and time
(in our scenario the nucleation of an isolated
bubble) which propagates at a speed asymptotically
approaching the speed of light, thus producing
smooth surfaces of constant negative spatial curvature.
By contrast, to single out surfaces of constant
positive curvature subsequent to the beginning of
inflation would seem to require some sort of
{\it nonlocal} process.

2. One possible source of error in our calculation
of the density perturbations is the neglect of
gravitational perturbations in region II. In region
II we used the ``stiff" approximation, in which the
effect of metric perturbations on the scalar field
perturbations is ignored. We know that the `stiff'
approximation works well while modes are well within
the Hubble radius. Therefore, for $\zeta \gtorder
1,$ in other words for modes well within the Hubble radius
at the beginning of the region I phase of inflation,
we do not expect using the stiff approximation in region
II to be a significant source of error. However, modes
with $\zeta \ll 1$ are never within the Hubble radius
at the beginning of region I.
This is because as one approaches the beginning
of region I, formally at least, the universe
becomes curvature dominated, and the size of
the co-moving Hubble radius
approaches a constant. Fortunately, since consistency
with observation requires that $\Omega \ge 0.1$--$0.2,$
all scales accessible to observation (i.e., within
the present apparent horizon) are never far from the range of
validity for the stiff approximation in region II.

3. In calculating a density perturbation spectrum,
for purposes of computational simplicity, we assumed
that the effective mass of the inflaton field
changes instantaneously from $m^2=2 H^2$ outside the
bubble wall to $m^2=0$ inside the bubble wall,
which is assumed to be infinitely thin. One would
expect a slightly different spectrum for a wall
of finite thickness, especially for large wave
numbers. In particular for $\zeta $ large in relation to
the inverse of the wall thickness, the modes
respond to the change in mass adiabatically rather
than according to the sudden impulse approximation.
We have extended our calculation to the
case $m^2 \gg 2H^2$ case, in which case the `thin wall'
approximation would be fully justified. The results
will be presented elsewhere.\refmark{\btnew } On scales
accessible to observation it turns out that varying
$m^2/H^2$ has little effect on the power spectrum.

4. Refs. \lyth ~ and  \rp ~ consider
a scenario of open inflation restricted to region I.
As an initial condition they impose the requirement
that the initial state is annihilated by the operators
associated with modes that have positive frequency with
respect to conformal time (i.e., with the asymptotic
behavior $t^{-i\vert \zeta \vert -1}$ near the light cone).
This choice of initial condition is not connected to
what happens outside the light cone (prior to the
coordinate singularity at $t=0$) and not surprisingly
gives unphysical behavior near the light cone: It is a
state akin to the Rindler vacuum for ordinary Minkowski
space. For such a state the stress-energy observed by
a freely falling observer crossing the light cone diverges;
for the Bunch-Davies vacuum there is no such divergence.

Except for the factor of $\coth [\pi \xi ]$ in eqn. \pspchi ,
our power spectrum is in agreement with refs. \lyth ~
and \rp . This is precisely the expected discrepancy,
resulting from the Bogolubov transformation, because
as far as the evolution in region I is concerned, we are
in agreement with  refs. \lyth ~ and \rp .

5. In the model of open inflation scenario presented here
the ``big bang" singularity at $t=0$ in the open expanding
FRW universe is not a genuine singularity. Rather it
has been reduced to a coordinate singularity, similar
in character to the coordinate singularity of the black
hole horizon in conventional Schwarzschild coordinates.
A freely falling observer passing from region II into
region I would not experience any singularity. With
this `big bang' singularity removed, it is tempting
to contemplate a universe eternal in the backward time
direction, with no initial singularity or Planck era.
Unfortunately, as shown by Vilenkin
and by Borde and Vilenkin, \refmark{\bv }
eternal inflation backward in time seems to be
 inconsistent with a
finite bubble nucleation rate.

6. We did not consider the generation of gravitational
waves in our model. One would expect a sizable
contribution to the CMB from tensor modes when
the energy scale of inflation is close to the
Planck scale, just as in flat inflation.

{\bf Acknowledgements:} We would like to thank Bruce Allen,
Robert Brandenberger,
Robert Caldwell, Edmund Copeland, Andrew Liddle, David Lyth,
Jim Peebles, Bharat Ratra, Martin Rees, Misao Sasaki,
and Frank Wilczek for useful discussions. Our result for the
Bogolubov coefficients in eqn. \jjy ~ has been obtained independently
by Allen and Caldwell \refmark{\allencaldwell} and
by Yamamoto, Tanaka and Sasaki\refmark{\sasakitanaka} using different
techniques and assumptions. We thank them for showing us
their calculations prior to publication.
MB was partially supported by the U.S. Department of
Energy under contract DE-F602-90-ER40542.
The work of MB and NT was  partially supported by
NSF contract PHY90-21984 and by the David and Lucile Packard
Foundation. We would also like to thank the Isaac Newton
Institute for Mathematical Sciences in Cambridge for their
hospitality. The work of ASG was supported in part by the
National Science Foundation under grant PHY93-09888.

\centerline{\bf Note Added}

We have recently extended the calculation of the
spectrum of density perturbations to the case
of arbitrary false vacuum mass $m$ (assumed positive).
The result is surprisingly
insensitive to the precise value of $m$, changing
little as $m$ varies from zero to infinity.
We have also received a recent paper by Yamamoto et al.
\refmark{\yamnew}
in which they attempt to perform the same calculation as
we have done here using instead analytic continuation of the
Euclidean vacuum modes, with
no change in the mass of the scalar field occurring
across the bubble wall (they assume
$m^2$ is constant and much less than
$H^2$ everywhere in de Sitter space).
If we make this assumption (which seems hard to justify
physically)
in our approach, our result is different from theirs.
We attribute this to the
failure of the Euclidean continuation method
to correctly describe the matching of quantum field
modes across the bubble wall
(this is discussed in detail in ref. \btnew).

\refout

\APPENDIX{A}{A---Matching Conditions for Minkowski Space.}

In this appendix we present the details of the derivation
of the matching conditions for Minkowski space for the
expansions in regions I and II assuming a massless scalar
field. The calculation is carried out only for the s-wave.

Our strategy is the following. We expand $\tilde \phi $
in terms of the more customary mode expansion
$$\eqalign{
\tilde \phi (r_m, t_m)&=\int _0^{\infty }dk~k^2j_0(kr_m)
\times \left\{ a^{(s)}(k)\cos [kt_m] +
a^{(a)}(k)\sin [kt_m]\right\} ,\cr } \eqn\appaa$$
which is nonsingular on the light cone and valid in both
regions I and II.

The expansion in region II
$$\eqalign{
\tilde \phi (\sigma ,\tau )&=\int _{-\infty }^{+\infty }
d\zd ~\sigma ^{+i\zd -1}
\times \left\{ A^{(s)}_{II}(\zd ) Q^{(s)}(\tau ;\zd )
+A^{(a)}_{II}(\zd ) Q^{(a)}(\tau ;\zd ) \right\} \cr
&=\int _{-\infty }^{+\infty }
d\zd {\sigma ^{+i\zd -1}\over \cosh [\tau ]}
\times \left\{ A^{(s)}_{II}(\zd ) \cos (\zd \tau )
+A^{(a)}_{II}(\zd ) \sin (\zd \tau ) \right\} \cr }\eqn\appab$$
contains enough information to determine the coefficient
functions $a^{(s)}(k)$ and  $a^{(a)}(k),$ because region II
contains a Cauchy surface, defined by $\tau =0.$

It follows that
$$\eqalign{
&\pmatrix{
a^{(s)}(k)\cr
a^{(a)}(k)\cr }=
{2\over \pi }\int _0^{\infty }r_m^2dr_m~ j_0(kr_m)
\times
\pmatrix{
\phi (r_m, t_m=0)\cr
{(1/k)}{\partial \over \partial t_m}\phi (r_m, t_m=0)\cr }
\cr
&~~~~={2\over \pi }\int _{-\infty }^{+\infty }d\zh
\int _0^{\infty }\sigma ^2d\sigma ~ {\sin [k\sigma ]\over k\sigma }
\cdot \sigma ^{+i\zd -1}\times
\pmatrix{
{}~A^{(s)}_{II}(\zd )\cr
{(\zd /k\sigma )}\cdot A^{(a)}_{II}(\zd )\cr }
\cr
=&\int _{-\infty }^{+\infty }d\zh
\pmatrix{ M^{(s)}(k;\zd )&0\cr 0& M^{(a)}(k;\zd )\cr }
\cdot \pmatrix{
A^{(s)}_{II}(\zd )\cr
A^{(a)}_{II}(\zd )\cr }\cr
}\eqn\appac$$
where
$$\eqalign{M^{(s)}(k;\zd )&=
{2\over \pi k}\int _0^\infty d\sigma \sin [k\sigma ]~
\sigma ^{+i\zd }\cr
&={2\over \pi }~\Gamma (1+i\zd )~
k^{-i\zd -2}\cosh \left( {\pi \zd \over 2}\right) ,
\cr }
\eqn\appad$$
and similarly
$$\eqalign{M^{(a)}(k;\zd )&=
{2\zd \over \pi k^2}\int _0^\infty d\sigma \sin [k\sigma ]~
\sigma ^{+i\zd -1}\cr
&={2i\zd \over \pi }~\Gamma (+i\zd )~
k^{-i\zd -2}\sinh \left( {\pi \zd \over 2}\right) ,
\cr }
\eqn\appae$$
where we have used the relation
$$\eqalign{& \int _0^\infty dt~t^\alpha \sin (at)=
a^{-1-\alpha }~~\Gamma (1+\alpha )~~
\cosh \left( {i\pi \alpha \over 2}\right) .\cr }\eqn\appaf$$

We now compute the coefficients of the region I expansion
$$\eqalign{\tilde \phi (\xi , t)&=
\int _0^\infty d\zh ~
{\sin [\zh \xi ]\over \sinh [\xi ]}
\times \left\{
A^{(+)}_I(\zh )~t^{+i\zh -1} +
A^{(-) }_I(\zh )~t^{-i\zh -1}
\right\}  ,\cr
}\eqn\appag$$
in terms of $a^{(s)}(k)$ and $a^{(a)}(k).$

The relation
$$\eqalign{ & \pmatrix{
A^{(+)}_I(\zh )\cr
A^{(-)}_I(\zh )\cr } =
{i\over \zh \pi }\int _0^\infty d\xi ~\sinh ^2\xi ~
{\sin [\zh \xi ]\over \sinh [\xi ]} ~\cr
&~~~~\times
\pmatrix{
[-i\zh -1]~t^{-i\zh +1}
&-t^{-i\zh +2}\cr
[-i\zh +1]~t^{+i\zh +1}
&t^{+i\zh +2}\cr }
\cdot
\pmatrix{ \tilde \phi (\xi , t)\cr
{\partial \over \partial t}\tilde \phi (\xi , t )\cr },
\cr }\eqn\appah$$
where $t>0$ is arbitrary but fixed,
may be rewritten as
$$\eqalign{
A^{(\pm )}_I(\zh )
&=
{i\over \zh \pi }\int _0^\infty d\xi ~\sinh ^2\xi ~ R_0(\xi ;\zh )~\cr
&~~~~\times
t^{\mp i\zh +1}\left[  (-i\zh \mp 1)~\mp t{\partial \over
\partial t}\right]
\phi (\xi ,t)\cr
&=t^{\mp i\zh +1}\left[  (-i\zh \mp 1)~\mp t{\partial \over
\partial t}\right] \cr
&~~~~\times
{i\over \zh \pi }\int _0^\infty d\xi ~\sinh ^2\xi ~
{\sin [\zh \xi ]\over \sinh [\xi ]}
\int _0^\infty k^2dk~ {\sin \bigl( kt\sinh \xi \bigr) \over
kt\sinh [\xi ]}\cr
&~~~~\times
\left\{
a^{(s)}(k)~\cos \bigl( kt\cosh \xi \bigr)
+a^{(a)}(k)~\sin \bigl( kt\cosh \xi \bigr)
\right\}  \cr
}\eqn\appai$$
Taking into account eqns. \appac --\appae , we may write
$$\eqalign{
A^{(\pm )}_I&(\zh )
=t^{\mp i\zh +1}\left[  (-i\zh \mp 1)~\mp t{\partial \over
\partial t}\right] \times
{i\over \zh \pi }\int _0^\infty d\xi ~\sinh ^2\xi ~
{\sin [\zh \xi ]\over \sinh [\xi ]}\cr
&~~~~\times
\int _0^\infty k^2dk~ {\sin \bigl( kt\sinh \xi \bigr) \over
kt\sinh [\xi ]}\int _{-\infty }^{+\infty }d\zd ~
{2\over \pi }\Gamma (i\zd )~k^{-i\zd -2}\cr
&~~~\times \Biggl\{
(i\zd )\cosh \left( {\zd \pi \over 2} \right)
\cos \bigl[ kt\cosh \xi \bigr] ~ A^{(s)}_{II}(\zd ) \cr
&~~~~~~~~+(i\zd )\sinh \left( {\zd \pi \over 2} \right)
\sin \bigl[ kt\cosh \xi \bigr]  ~
A^{(a)}_{II}(\zd )\Biggr\}  \cr
&=\int _{-\infty }^{+\infty }d\zd ~\left\{
M^{(\pm )}(\zh  ; \zd ,s)~A^{(s)}(\zd )+
M^{(\pm )}(\zh  ; \zd ,a)~A^{(a)}(\zd )\right\}
\cr }\eqn\appaj$$
We evaluate
$$\eqalign{
M&^{(\pm )}(\zh; \zd , s)=
{-2\zd \over \pi ^2\zh }\Gamma (i\zd )~
\cosh \left( {\zd \pi \over 2} \right) \cr
&\times t^{\mp i\zh +1}\left[  (-i\zh \mp 1)~\mp t{\partial \over
\partial t}\right] ~ t^{-1}
\int _0^\infty d\xi ~\sin [\zh \xi ]\cr
&\times \int _0^\infty dk~k^{-i\zd -1}
\sin [kt \sinh \xi ]
\cos [kt \cosh \xi ],\cr }\eqn\appak$$
first carrying out the integration over $k$
$$\eqalign{
\int _0^\infty dk &~k^{-i\zd -1}
\sin (kt\sinh \xi ) \cos (kt\cosh \xi )\cr
&={1\over 2}\int _0^\infty dk ~k^{-i\zd -1}
\left[
\sin (kt~e^\xi )-
\sin (kt~e^{-\xi })
\right] \cr
&={1\over 2}~\Gamma (-i\zd )~\cosh \left[ {i\pi \over 2}
(-1-i\zd )\right] \times
\left[ (te^\xi )^{i\zd } -(te^{-\xi })^{i\zd }
\right] \cr
&=\Gamma (-i\zd )~\sinh \left( {\zd \pi \over 2}
\right) ~~t^{i\zd }~~\sin (\zd \xi ),\cr
}\eqn\appal$$
and then the integration over $\xi $
$$\eqalign{
\int _0^\infty d\xi ~&\sin [\zh \xi ]\sin [\zd \xi ]
={\pi \over 2}\cdot
\biggl\{  \delta \bigl( \zh -\zd \bigr)
-\delta \bigl( \zh +\zd \bigr) \biggr\}  ,\cr } \eqn\appam$$
one finally obtains
$$\eqalign{
M^{(\pm )}(\zh ;\zd , s)&=
t^{\mp i\zh +1}\left[  (-i\zh \mp 1)~\mp t{\partial \over
\partial t}\right] t^{+i\zd -1}\cr
&\times {-\zd \over \zh \pi }~ \Gamma (i\zd )~\Gamma (-i\zd )
\sinh \left( {\zd \pi \over 2}\right) ~
\cosh \left( {\zd \pi \over 2}\right) \cr
&\times \biggl\{ \delta (\zh -\zd )
-\delta (\zh +\zd ) \biggr\} .\cr }\eqn\appan$$
Note that
$$\eqalign{&
t^{\mp i\zh +1}\left[ (-i\zh \mp 1)~\mp t{\partial \over
\partial t}\right] t^{+i\zd -1}\cr
&~~~=\mp i(\zd \pm \zh )~t^{+i(\zd \mp \zh )}\cr }
\eqn\appao$$
vanishes where one of the two delta functions
$\delta (\zh \pm \zd )$ is nonzero; therefore,
eqn. \appan ~ becomes
$$\eqalign{
M^{(\pm )}(\zh ;\zd , s)&=
{\zd \over \zh \pi }~ \Gamma (i\zd )~\Gamma (-i\zd )
\sinh (\zd \pi )~
\delta (\zd \mp \zd )\cr
&=~~{i\zd \over \zh }
\cdot \delta \bigl(\zd -(\pm \zd ) \bigr) .\cr
&=(\pm i)\cdot \delta \bigl(\zd -(\pm \zd ) \bigr) .\cr }
\eqn\appap$$
{}From the well-known relation
$\Gamma (x)\Gamma (1-x)=\pi /\sin (\pi x),$ it follows that
$\Gamma (ix)\Gamma (-ix)=\pi /(x\sinh [\pi x]),$ hence the
last line.

Similarly, we evaluate
$$\eqalign{
M&^{(\pm )}(\zh; \zd , a)=
{2i\over \zh \pi ^2}~\Gamma (i\zd )~(i\zd )~
\sinh \left[ {\zd \pi \over 2} \right] \cr
&\times t^{\mp i\zh +1}\left[  (-i\zh \mp 1)~\mp t{\partial \over
\partial t}\right] ~ t^{-1}
\int _0^\infty d\xi ~\sin [\zh \xi ]\cr
&\times \int _0^\infty dk~k^{-i\zd -1}
\sin [kt \sinh \xi ]~
\sin [kt \cosh \xi ].\cr }\eqn\appaq$$
Integrating over $k,$ we obtain
$$\eqalign{
\int _0^\infty &dk ~ k^{-i\zd -1}~~
\sin (kt \sinh \xi )
\sin (kt \cosh \xi )\cr
&={1\over 2}
\int _0^\infty dk ~ k^{- i\zd -1}~~\biggl[
\cos (kte^{-\xi })- \cos (kte^{+\xi }) \biggr] \cr
&={1\over 2}~\Gamma (-i\zd )~\cosh \left( {\pi \zd \over 2}\right)~
\left[
(te^{-\xi })^{+i\zd }- (te^{+\xi })^{+i\zd }
\right]\cr
&=(-i)~
\Gamma (-i\zd )~\cosh \left( {\pi \zd \over 2}\right)~
t^{+i\zd }~~\sin (\zd \xi ),\cr
}\eqn\appar$$
so that we have
$$\eqalign{
M^{(\pm )}(\zh; \zd , a)&=
{2i\zd \over \zh \pi ^2} ~\Gamma (-i\zd )~\Gamma (+i\zd )~
\cosh \left( {\pi \zd \over 2}\right)~
\sinh \left( {\pi \zd \over 2}\right)\cr
&\times t^{\mp i\zh +1}\left[  (-i\zh \mp 1)~\mp t{\partial \over
\partial t}\right] ~ t^{+i\zd -1}\cr
&\times {\pi \over 2}~
\biggl\{ \delta (\zh -\zd ) -\delta (\zh +\zd ) \biggr\} \cr
&={\zd \over \zh }\delta \biggl( \zd -(\pm \zh )\biggr) \cr
&=(\pm )\cdot \delta ( \zd -[\pm \zh ]).\cr }
\eqn\appas$$
Therefore, for $\zeta >0,$
$$\eqalign{
A^{(+)}_I(\zeta )&=+[~+iA^{(s)}_{II}(+\zeta )+A^{(a)}_{II}(+\zeta )],\cr
A^{(-)}_I(\zeta )&=-[~+iA^{(s)}_{II}(-\zeta )+A^{(a)}_{II}(-\zeta )].\cr
}\eqn\appat$$

We may summarize this result in a more compact and more
intuitive way as follows. In terms of the mode functions,
as one passes from region II to region I, one has (for
$\zeta >0$)
$$\eqalign{
\sigma ^{+i\zeta -1}\cdot
{e^{+i\zeta \tau }\over \cosh [\tau ]}
&\to
(2i)\cdot {\sin [\zeta \xi ]\over \sinh[\xi ]}
\cdot t^{+i\zeta -1},\cr
\sigma ^{+i\zeta -1}\cdot
{e^{-i\zeta \tau }\over \cosh [\tau ]}
&\to 0,\cr
\sigma ^{-i\zeta -1}\cdot
{e^{+i\zeta \tau }\over \cosh [\tau ]}
&\to 0,\cr
\sigma ^{-i\zeta -1}\cdot
{e^{-i\zeta \tau }\over \cosh [\tau ]}
&\to
(-2i)\cdot {\sin [\zeta \xi ]\over \sinh[\xi ]}
\cdot t^{-i\zeta -1}.\cr
}\eqn\svea$$
We finally point out that these matching conditions are
generally applicable to curved space and nonvanishing
mass. It is the behavior of the leading singularities at
the light cone that determine the matching conditions;
therefore, introducing mass and spacetime curvature gives
subdominant effects.

\vfill
\eject

\leftline{\bf Figure Captions}

\item{Fig. ~1}{ $SO(3,1)$ symmetric coordinates. Spacetime
is divided into three regions---regions I, II, and III.
In the zeroth order solution the scalar field is constant
on each of the hyperboloids, and on
the light cone of the origin $O$. The dashed line shows the origin of
spherical region I coordinates ($\xi =0$).
}

\item{Fig. ~2}{Slice through a  bubble.
 In the (lower) Euclidean
domain a bubble nucleates via the Coleman--De Luccia instanton.
 In the
(upper) Lorentzian domain the bubble expands classically at
a speed approaching the speed of light.
Region I, the bubble interior, is an
expanding FRW universe with spatial hypersurfaces of
constant negative curvature.
Region II is the exterior of the expanding bubble.
The scalar field, and background density are
constant along curves shown. The future null cone of the
center of the nucleating bubble is shown as
the line $r=t$. The background
scalar field takes
the value $\phi_n$ on this null surface.}

\item{Fig. ~3}{Potential for `Old'+`New' Inflation. Initially
the inflaton field is stuck in the false vacuum F, during the
epoch of old inflation. This is
exited by the nucleation of a single bubble, which eventually
grows to encompass our entire observable universe. Instead
of tunneling to the true vacuum T, the scalar field tunnels
onto a slow rollover potential, at the point
$\phi_n$, and the interior of the
bubble expands quasi-exponentially, during the epoch of
new inflation. In our scenario, this is
shortened compared to conventional inflation
so that the negative spatial curvature survives.
Then as the field begins to roll more rapidly, and starts to
oscillate about the true vacuum value
$\phi_r$, reheating occurs,
 converting vacuum energy into radiation and matter.
}

\end